\documentclass[aps,superscriptaddress]{revtex4}
\usepackage{amssymb,amsmath,epsfig}
\usepackage[colorlinks=true, pdfstartview=FitV, linkcolor=blue, citecolor=red, urlcolor=magenta, breaklinks=true]{hyperref}
\usepackage{graphicx}
\usepackage{color}
\usepackage{epstopdf}
\usepackage[capposition=top]{floatrow}
\usepackage{subfig}

\begin{document}
\title{Phenomenology of Strong Interactions --- Towards an Effective Theory for Low Energy QCD}

%twocolumn[  \begin{@twocolumnfalse}
\author{Adamu Issifu}\email{ai@academico.ufpb.br}
\affiliation{Departamento de F\'isica, CFM---Universidade Federal de Santa Catarina, \\
 Caixa Postal 476, 88040-900 Florian\'opolis, SC, Brazil}
\affiliation{Departamento de F\'isica, Universidade Federal da Para\'iba, 
Caixa Postal 5008, 58051-970 Jo\~ao Pessoa, Para\'iba, Brazil}

\author{Francisco A. Brito}\email{fabrito@df.ufcg.edu.br}
\affiliation{Departamento de F\'isica, Universidade Federal da Para\'iba, 
Caixa Postal 5008, 58051-970 Jo\~ao Pessoa, Para\'iba, Brazil}
\affiliation{Departamento de F\'{\i}sica, Universidade Federal de Campina Grande
Caixa Postal 10071, 58429-900 Campina Grande, Para\'{\i}ba, Brazil}

\begin{abstract}
In this paper, we develop models applicable to phenomenological particle physics by using the string analogy of particles. These theories can be used to investigate the phenomenology of confinement, deconfinement, chiral condensate, QGP phase transitions, and even the evolution of the early universe. Other confining properties such as scalar glueball mass, gluon mass, glueball-meson mixing states, QCD vacuum, and color superconductivity can also be investigated in these model frameworks. We use one of the models to describe the phenomenon of color confinement among glueballs at the end of the paper. The models are built based on the Dirac-Born-Infeld (DBI) action modified for open strings with their endpoints on a D$p$-brane or brane-anti-brane at a tachyonic vacuum. %At the tachyonic vacuum the strings are connected by a flux line on the D-brans forming closed strings. Also, the strings on the brane can be adjusted without any cost of energy. Furthermore, we couple the theory with Dirac's Lagrangian for free particles following the appropriate gauge invariant principles. We systematically modify all the color neutral fields in Dirac's Lagrangian with a color dielectric function, $G(\phi)$, which is obtained naturally from the modified DBI action. This enable us to use Abelian gauge theory in place of non-Abelian gauge theory to investigate QCD properties. Finally, we show that the color dielectric function vanishes when we consider a non-Abelian gauge by deriving the Yang-Mills theory.

%As an example, we will show how one of the theories reproduces some features of  confinement phenomenology and its associated properties. 

\end{abstract}
\maketitle
\pretolerance10000
\section{Introduction}
String theory was conceived in the late 1960s to provide an explanation for the behavior of nuclear matter such as protons and neutrons \cite{Schwarz, Schwarz1}. Even though the theory was not successful in explaining the characteristics of quarks and gluons at its inception, it promised an interesting intervention in other areas of physics \cite{Green, Polchinski2, Johnson, Zwiebach1, Becker}. It showed capability in giving an insight into cosmology, astrophysics, and unification of the fundamental forces of nature which has been on the table of physicists for some time now. Consequently, the theory of Quantum Chromodynamics (QCD) was developed in the early 1970s to give a comprehensive explanation of nuclear matter \cite{Gross, Politzer}. The QCD theory is now accepted as the standard theory for strong interactions. Interestingly, recent development shows that string theory and QCD describe the same physics.

The formulation of Quantum Electrodynamics (QED) and QCD are almost the same. They are both formulated from field theory based on gauge theory which forms the foundation of the highly acceptable Standard Model (SM). The fundamental particles of these theories are studied based on the gauge boson they interchange: Photons for electrodynamic force, $W^{\pm}$ and $Z^0$ bosons for electroweak force and gluon for strong nuclear force \cite{Quigg, Collins1, Shaw}. Subsequently, gravitational force does not fall under this category, so they are investigated under Einstein's general theory of relativity. Physicists have bunged their hope on string theory to unify all the fundamental forces, this will fall under 'physics beyond the SM' \cite{Aharony, Maldacena, Grana, Polchinski3, Haro}. Upon the similarities, quarks and gluons are color particles classified under three conventional colors (red, blue, and green) whilst photons are color neural bosons that mediate electrically charged leptons. Also, gluons self-interact due to their color charges but photons do not \cite{Quigg, Collins1}. 

Additionally, QCD falls short in explaining color-neutral particles such as bound states of gluons (glueballs) and quarks (hadrons and mesons), so string theory can be resorted for further description. The string-like description of hadrons arising from the quark model \cite{Amsler} is an important phenomenon in applying string theory. Under this picture, when a quark and an antiquark are pulled apart, they behave as if they are connected by a rubber band (gluon) which becomes increasingly difficult to separate when the separation distance between them keeps increasing. This analogy gradually fails when the particles are brought closer and closer together. In this regime string theory fails and QCD theory becomes viable. Now, looking at the particles in terms of fundamental strings, string theory describes hadrons quite well and provides the background for the unification of the fundamental forces. In string theory, the strings are treated as particles, where different particles are associated with different string oscillations. The masses of the  particles are also associated with the energy of the oscillating string. The intrinsic spin of the particle is associated with the two perpendicular oscillations of the string with its endpoints fixed on D-branes, similar to the direction of electric and magnetic fields in a photon. Hence, photons and gluons can be identified in terms of the open strings as spin-1 particles. The clockwise and anti-clockwise movement of the closed strings makes it possible to classify it among spin-2 bosons such as graviton. Since, QCD involves color fields, to study it under string theory we assume that the endpoints of the open strings on a D-brane serve as the source and sink of the color charge \cite{Bigazzi}.

It has been conjectured that open bosonic strings studied at a tachyonic vacuum behave as if they are closed strings with no D-branes. However, soliton solutions in this region point to the presence of lower dimensional branes \cite{Sent2, Bardakci}. This projection is also corroborated in superstring theories \cite{Sent1, Sen1} and evident in the first \cite{Recknagel, Sen2, Harvey} and second \cite{Kostelecky, Sen, Zwiebach2, Kostelecky1, Berkovits, Harvey1, Gopakumar, Gerasimov} quantizations in string theories \cite{Sent4}. At the tachyonic vacuum, the negative energy density $V(T_0)$ of the tachyons in the vacuum exactly cancels \cite{Witten1} the energy density of the D-branes $\varepsilon_p$ i.e. $V(T_0)+\varepsilon_p=0$. In non-BPS D-branes $\varepsilon_p=\tau_p$ where $\tau_p$ is the D-brane tension \cite{Lindstrom,Gustafsson,Sent3}. {It should be noted that at $|T| = T_0$ the total energy density of the tachyons and the brane tension vanishes identically, signifying the absence of open strings at that point.} In this view, at a tachyonic vacuum, there will be no physical open string excitations because there are no D-branes. On the contrary, the 'usual' field theory gives an alternative explanation, because shifting the vacuum and expanding around its 'true' minimum can change a negative square mass of the tachyons to a positive one even though it does not completely remove all the states. Since it will not cost any energy to adjust the fundamental strings on the worldvolme of the D-branes at a tachyonic vacuum, it will be difficult to notice their presence. Hence, fluctuations around the vacuum represent the presence of lower dimensional D-branes \cite{Sent2, Bardakci}. %For bosonic open strings with tachyonic modes like the ones intended for this paper, the tachyons will have negative energy density at its extremum $V(0)$ which cancels out the tension of the D-branes. Consequently, the extremum represents bosonic closed strings in the vacuum with their endpoints on D-branes \cite{Berkovits}. 
These phenomena have been investigated in references \cite{Sen, Zwiebach2, Kostelecky} using open string field theory \cite{Witten2}.

In this paper, we modify the Dirac-Born-Infeld (DBI) action, so that the associated open strings with their endpoints on the remaining lower-dimensional branes at the tachyonic vacuum can be analyzed. The objective is to develop models that can mimic QCD theory both in the UV and IR regions with UV safety. %In the process we develop theories that mimic QCD theory in quench and unquench approximations. 
The models can be applied in developing potential models such as the linear confining and Cornell potential models. In the analysis, we consider that the string worldsheet falls inside the D-brane worldvolume making it easy for the endpoints of the strings to be connected by the flux line on the worldvolume to form a close string suitable for modeling {\it color confinement} in a flux-tube like picture. The dynamics of the strings tangential to the D-brane worldvolume is represented by the gauge field $F_{\mu\nu}$ and the component transverse to the worldvolume is represented by a massless scalar field $X^a$. The net flux involved is determined by the source and sink of the flux carried by the endpoints of the string on the lower dimensional remnants of the original D-brane in the vacuum. Also, the condition for minimum energy warrants that the flux does not spread because the source and sink of the flux emanate from a pointlike object on the D-brane worldvolume. %The outcome is a fundamental string extended along a compact direction.  

The paper is organized as follows: In Sec.~\ref{RTD} we review Tachyons and D-branes, divided into two subsections, under Sec.~\ref{T} we review Tachyons, and in Sec.~\ref{D} we review D-branes. We present the Modification of the Dirac-Born-Infeld Action which serves as the bases for the study in Sec.~\ref{mdbia}. Also, in subsection \ref{DRYMT} we review the Dimensional Reduced $\text{U}(1)$ Yang-Mills theory and its intended consequences in relation to the Standard Model (SM) of Particle Physics. Section \ref{bstv} and subsequent sections contain our original contributions to the subject. We present the Bosonic String at Tachyonic Vacuum in Sec.~\ref{bstv}, where we present the details of dimensional reduction from $10D$ Dirac-Born-Infeld action to $4D$ conducive for describing SM particles. We present Gauge Theories Modified with $G(\phi)$, in Sec.~\ref{GTM}, this section was divided into two. We studied Fermions Coupled to Fundamental Strings at Tachyonic Vacuum in Sec.~\ref{mts} and Non-Abelian Gauge Theory in Sec.~\ref{NAGT}. Section \ref{TM} contains the Phenomenon of Gluon Confinement, divided into four subsections. We present The Model in Sec.~\ref{TMA}, Confining Potentials in Sec.~\ref{CP}, Gluon Condensation in Sec.~\ref{GC} and strong Running Coupling and QCD $\beta$-function in Sec.~\ref{SRC}. we present our findings and Conclusions in Sec.~\ref{C}. 

%= + 0 / )
\section{Review of Tachyons and D$p$-branes}\label{RTD}
\subsection{Tachyons}\label{T}
{Generally, tachyons are classified as particles that travel faster than light or weakly interacting superluminal particles. Relativistically, single particle energy is expressed as $E^2=p^2c^2+m^2c^4$, where $p$ is the spatial momentum and $m$ is the mass of the particle. For a particle to be faster than light, the relativistic velocity $\beta=pc/E>1$. Hence, for tachyons with real $p$, $m$ must necessarily be imaginary \cite{Bilaniuk}. However, this analogy does not make a strong and convincing case for the tachyons. Rather, Quantum Field Theory (QFT) provides some satisfactory explanation to the dynamics of tachyons. QFT suggests that particles that travel faster than light does not exist in nature. So tachyons are simply unstable particles that decay. Based on this understanding, we consider a scalar field, say $\phi$, with the usual kinetic term and a potential $V(\phi)$ whose extremum is at the origin. If one carries out perturbation quantization about $\phi=0$ up to the quadratic term and ignores the higher order terms in the action, we obtain a particle-like state at $V''(0)$. These results have two interesting interpretations; if $V''(0)>0$ we have a particle of mass $m^2_\phi$ but for $V''(0)<0$ we have a tachyonic state with a negative $m^2_\phi$. In this case, tachyons can be given a physical meaning. So far, we know that the tachyons have negative $m^2_\phi$ and potential whose maximum is at the origin, thus a small displacement of $\phi$ at the origin will cause it to grow exponentially with time towards the true minimum. %Consequently, perturbation theory where we ignore the cubic and higher order terms breaks down. 
By the above description, tachyons can be represented by a potential such as 
\begin{equation}
V(\phi)=-\dfrac{1}{2}m_\phi^2\phi^2+ c_3\phi^3+c_4\phi^4\cdots,
\end{equation}
where $c_3\,\text{and}\,c_4$ are constants. Hence, tachyonic fields are associated with the 'usual' Higgs fields where the Higgs particles acquire negative square mass at the true minimum of their potential. %One way of removing the tachyonic mode under QFT is by expanding the potential about its true minimum. This changes the $-m^2_\phi$ to positive.  
%= + 0 / )
Accordingly, tachyons in QFT are related to instability that breaks down the perturbation theory in normal field theory. The usual quantization where the cubic and higher order terms are considered small corrections to the quadratic term is no longer tenable. Since $V(\phi)$ has its maximum at $\phi=0$, it renders it classically unstable at that point. So one cannot guarantee that the fluctuation at that point is small. This behavior comes with an inbuilt solution i.e. one can expand the potential about the true minima $\phi_0$ up to the quadratic term and proceed with the perturbation quantization about that point. Thus, the cubic and higher-order terms in the expansion can be discarded. This process will lead to the creation of a particle with a positive mass $m^2_\phi$ in the spectrum. This process removes the tachyonic modes in the spectrum. Additionally, in D$p$-brane systems, the theory must be invariant under $Z_2$ symmetry, i.e. $\phi\rightarrow-\phi$ and in the presence of brane-anti-brane, the theory must necessarily be invariant under phase symmetry $\phi\rightarrow e^{i\alpha}\phi$. %and also the field is small such that higher other terms can be ignored.
%= + 0 / )
Indeed, there are some benefits in working in tachyonic modes otherwise we can define a new field, $\eta=\phi-\phi_0$, and express the potential in terms of the new field,
\begin{equation}
V(\eta)=V(\eta+\phi_0).
\end{equation}
Working with this potential from the onset will remove the tachyonic mode from the spectrum because $V''(\eta=0)$ will be positive. However, there are some benefits to working with tachyonic fields, for instance, they possess high symmetry. This symmetry might not be explicit in $V(\eta)$ as it is in $V(\phi)$. The high symmetry in tachyonic fields leads to the phenomenon of spontaneous symmetry breaking, where the potential has more than one minima i.e. $V(\phi)=V(-\phi)$ corresponding to $\pm\phi_0$. This phenomenon is well known in elementary particle physics \cite{Sent3}.
%QFT provides second quantization of a particle however, string theory conventionally provides first quantization formalism. Under this formalism, the spectrum of a particle state in theory are observed through quantization of the vibrational modes of a string.
}
\subsection{D$p$-branes}\label{D}
Recent advances in string theory have provided some explanations regarding nonperturbative features of QCD theory, thanks to the discovery of D$p$-branes. The study of D$p$-branes gives insight into physically relevant systems such as {\it black holes}, supersymmetric gauge theories, and the connection between Yang-Mills theories and quantum gravity. Upon the numerous progress, a consistent nonperturbative background-independent construction of the theory is yet to be developed. This poses a challenge in directly addressing cosmological problems. %Besides, string theory holds the key to developing background-independent nonperturbative approach to string theory.
There are five known ways for which supersymmetric closed strings can be quantized, i.e. IIA, IIB, I, heterotic $\text{SO}(32)$ and heterotic $E_8\times E_8$ superstring theories. Individually, they give a perturbative description of quantum gravity. However, they are connected through duality symmetries \cite{Hull,Witten}. Indeed, some of these dualities are nonperturbative in nature because the string coupling $g_s$ in one theory may have an inverse relation $1/g_s$ with the other. The superstring theories together with M-theory (a theory that unifies superstring theories) consist of an extended object of higher dimensions, called D$p$-branes. Each of these theories contains branes but in different complements. Particularly, IIA/IIB superstring theories consist of even/odd D$p$-brane dimensional configurations. %The brane in one theory can be related to another theory through duality transformations. 
Using the appropriate duality transformations, one brane can be mapped onto the other even with the strings connecting them. Thus, none of the branes are seen to be more fundamental to others. This process is commonly referred to as 'brane democracy'.

Before moving into the specific kind of strings set out for this study, we will shed light on the features of open and closed strings.  Closed strings are topologically the same as a circle $S^1$, i.e. $[0,2\pi]$. They give rise to a massless set of spacetime fields, identified as graviton $g_{\mu\nu}$, dilaton $\varphi$, antisymmetric two-form $B_{\mu\nu}$ and an infinite set of massive fields upon quantization. Supersymmetric closed strings are also related to massless fields identified with graviton supermultiplets such as Ramond-Ramond $p$-form fields $A^{(p)}_{\mu,\cdots,\mu_p}$ and gravitino $\psi_{\mu\alpha}$. Consequently, the quantum theory of closed strings is naturally related to the theory of gravity in spacetime. On the other hand, open strings are topologically similar to the interval $[0,\pi]$. They give rise to massless gauge field $A_\mu$ in spacetime upon quantization. Supersymmetric open strings are also associated with massless gaugino $\psi_\alpha$. Accordingly, open strings  have their ends on Dirichlet $p$-branes (D$p$-brane) and the gauge field lives on the worldvolume of the D$p$-brane. Upon these differences, the physics of closed and open strings are related at the quantum level. Closed strings were first observed through the one-loop process of an open string. Under this process, close strings appeared as poles in nonplanar one-loop open string diagrams \cite{Gross1,Lovelace}. The open strings have their endpoints on the D$p$-branes whilst close strings have no endpoints. The degree of freedom of the open strings is associated with standing wave modes on the fields while the close strings correspond to left-moving and right-moving waves. The boundary conditions for open strings on the bosonic field $X^M$ are Neumann (freely moving endpoints) and Dirichlet boundary conditions (fixed endpoints). The close string on the other hand corresponds to periodic and anti-periodic boundary conditions.
%= + 0 / )

Some D$p$-branes have unstable configurations both in the supersymmetric or the nonsupersymmetric string theories. The instability is attributed to tachyonic modes with negative square mass $m^2_\phi\,<\,0$ in the open string spectrum, we are interested in investigating the effects of the tachyonic modes \cite{Sen,Zwiebach}. Some D$p$-brane configurations with open strings containing tachyons are:

{\hskip 2em}{\bf Brane-antibrane}; it is a type IIA or IIB string theory with parallel D$p$-branes separated by $d\,<\,l_s$ ($l_s$ is the string length scale). They also carry tachyons in their open string spectrum. The difference in their orientation leads to opposite Ramond-Ramond (R-R) charges \cite{Hughes}. So the brane and anti-brane pair can annihilate leaving a neutral vacuum because the net R-R charge will be zero.

{\hskip 2em}{\bf Wrong-dimension branes}; the D$p$-brane with wrong dimension for type IIA/IIB with odd/even spatial dimension for $p$ instead of even/odd dimensions carry no charges under classical IIA/IIB supergravity fields. They have tachyons in their open string spectrum. Such branes can annihilate to form a vacuum without violating charge conservation.

{\hskip 2em}{\bf Bosonic D$p$-branes}; just as the wrong-dimension branes of type IIA/IIB string theory, the D$p$-brane of any particular dimension in the bosonic string theory have no conserve charge and has tachyons in their open string spectrum. Also, they can annihilate to form a neutral vacuum without violating charge conservation.

Again, even though the non-BPS D$p$-branes of type IIA/IIB string theory are unstable owing to the presence of tachyonic modes on their worldvolume \cite{Sen1}. We can obtain stable non-BPS branes by taking orientifolds/orbifolds of the theory which projects out the tachyonic modes \cite{Sen1,Bergman,Sen2}. %This process leads to projecting out the subset of the massless degrees of freedom so, the D-brane worldvolume and the corresponding action of the D-branes can be derived by following an appropriate truncation of the unsuitable terms.
%Applying string theory to such studies is appropriate because
 
\section{Modification of Dirac-Born-Infeld Action}\label{mdbia}
We begin the study with the Born-Infeld action (BI) \cite{Born} sometimes referred to as Dirac-Born-Infeld action (DBI) \cite{Gibbons,Dirac}. The focus will be on D$p$-branes \cite{Polchinski,Taylora} which are nonperturbative states on which open strings live. They are equally coupled with closed strings, Ramond-Ramond states and other massive fields. The nonperturbative nature of the action makes it possible to describe low energy degrees of freedom of the D$p$-branes \cite{Leigh} making it possible for application in low energy QCD. The distinction between the DBI action, other $p$-branes and supermembrane theories \cite{Hughes,Bergshoeff,Townsend} is the presence of gauge field in the worldvolume of DBI. The gauge field is associated with virtual open string states. %with their ends fixed on the world volume of the branes. 
Consequently, we obtain confinement of the fundamental strings with their endpoints fixed on the branes. Generally, the action can be expressed as  
\begin{equation}\label{tt1}
S=-T_p\int d^{p+1}\xi e^{-\varphi}\sqrt{-\text{det}\left( G_{\mu\nu}+B_{\mu\nu}+(2\pi\alpha')F_{\mu\nu}\right)}+S_{CS}+\text{fermions},
\end{equation}
where $G_{\mu\nu}$, $B_{\mu\nu}$ and $\varphi$ are the induced metric tensor, antisymmetric tensor, and dilaton field to the D$p$-brane worldvolume respectively. %Generally, there is a fermion coupling $i(2\pi\alpha')^2\bar{\Theta}\Gamma_\mu\partial\nu\Theta$ which appears under the square root, but for now, we do the analyses without the fermions and incorporate their dynamics soon when appropriate. 
Also, $F_{\mu\nu}=\partial_\mu A_\nu-\partial_\nu A_\mu$ is the worldvolume electromagnetic field strength of $A_\mu$ and $S_{CS}$ is a set of Chern-Simon terms while 
\begin{equation}\label{tt2}
\tau_p=\dfrac{T_p}{g_s}=\dfrac{1}{g_s\sqrt{\alpha'}(2\pi\sqrt{\alpha'})^p},
\end{equation}
is the brane tension and $g_s=e^{\langle\varphi\rangle}$ is the string coupling, with associated string tension
\begin{equation}
T_{\text{string}}=\dfrac{1}{2\pi\alpha'}.
\end{equation}

In type IIA/IIB string theory with $p$ even/odd are associated with quantum open strings containing massless $A_\mu,\,\mu=0,1,\cdots,p$ and $X^M,\, M=p+1,\cdots,9$ fields. These fields are the consequence of the gauge field living on the hypersurface and $X^\mu(\xi)$ transverse excitations. The geometry of D$p$-brane is not flat, so we generally define the embedding $X^\mu(\xi)$, where $\xi^\alpha$ represent $p+1$ coordinates on the D$p$-brane worldvolume $\sum_{(p+1)}$, and $X^\mu$ is the ten functions mapping from $\sum_{(p+1)}$ onto the spacetime manifold $\mathbb{R}^{9,1}$. 
 % c + 0 / ) =
%A specific limit of the Born-Infeld action in Eq.(\ref{tt1}) is required to study the low-energy aspect of the D-brane properties. %Again, where $X^a$ is a scalar field describing small transverse fluctuations of the brane around a flat hypersurface, when we assume that the D-branes are close to the hypersurface, we approximate $X^a=0$, $a\,>\,p$. 
Introducing the scalar field into the action, it becomes invariant under diffeomorphism and Abelian gauge transformations. So, a way of fixing the freedom of the former is to adopt a 'static gauge' such that
\begin{align}\label{tt2a}
X^\mu\equiv\xi^\mu \qquad\text{for}\qquad 0\,\leq\,\mu\leq p.
\end{align}
The remaining fields are
\begin{equation}\label{tt2b}
X^\mu\equiv X^M \qquad\text{for}\qquad (p+1)\,\leq\,M\,\leq d-p,
\end{equation}
which are the transverse coordinates to the worldvolume \cite{Tseytlin,Taylor,Taylor1} where $d$ is a spatial dimension. Thus, one can choose $d=9$ for superstring theory and $d=25$ for bosonic string theory. Under this gauge, the originally $d+1$-dimensional global poincar\'e symmetry spontaneously breaks down to a product of $p+1$ dimensional poincar\'e group with $d-p$ dimensional rotational symmetry group i.e. $\text{SO}(1,d)\rightarrow\text{SO}(1,p)\times\text{SO}(d-p)$. For a D$p$-brane possessing $p<d$ extends over $p$-dimensional subspace of $d$-dimensional space. The focus will be on a D$p$-branes with $p$-dimensional {\it hyperplanes} in $d$-dimensional space. Again, according to the static gauge fixed above, there are two possible consistent truncations;
\begin{itemize}
\item $X^M=0$; corresponds to pure BI theory \cite{Born} in $\mathbb{E}^{p,1}$ and 
\item $F_{\mu\nu}=0$; also corresponds to Dirac's theory \cite{Dirac} of minimal timelike submanifolds of $\mathbb{E}^{d,1}$.
\end{itemize}  %We will focus on simple Dp-branes: those that are p-dimensional hyperplanes inside the d-dimensional space. How can we specify such hyperplanes? We need (d − p) linear conditions. In three spatial dimensions (d = 3), a 2-brane (p = 2) is a plane, and it is specified by one linear condition (d − p = 3 − 2 = 1).
We can introduce the transverse scalar fluctuations by defining the induced metric as
\begin{equation}\label{tt3}
G_{\mu\nu}\approx\eta_{\mu\nu}+\eta_{MN}\partial_\mu X^M\partial_\nu X^N,
\end{equation}
that will approximate the D$p$-brane worldvolume to near flat. In this study, we are interested in understanding the dynamics of bosonic open strings with tachyonic modes in their spectrum, so we will proceed systematically toward that objective. The worldvolume theory of non-BPS D$p$-brane in IIA/IIB string theory corresponds to a massless $\text{U}(1)$ vector field, transverse oscillating scalar field $X^M$ and a tachyonic field $T$ \cite{Sent}. Accordingly, the leading order of the action results in dimensional reduction $10$-dimensional $\text{U}(1)$ Yang-Mills theory. Besides, higher-order corrections, $\alpha'=l_s^2$, to the order of the string scale  are also possible. Since we are proceeding with the assumption that the massless fields are slowly varying compared to the string length $l_s$, i.e. we discard the higher order derivatives and write the DBI action \cite{Leigh,Garousi,Callan} in a simple form without the explicit presence of the tachyons. Therefore, Eq.(\ref{tt1}) takes the form
\begin{equation}\label{tt1a}
S=-T_p\int d^{p+1}\xi e^{-\varphi}\sqrt{-\text{det}\left(\eta_{\mu\nu}+\eta_{MN}\partial_\mu X^M\partial_\nu X^N+B_{\mu\nu}+(2\pi\alpha')F_{\mu\nu}\right)}.
\end{equation}
Now we will express the extended form of the DBI by including the dynamics of the tachyons as studied in \cite{Garousi}. Accordingly,
\begin{equation}\label{tt1aa}
S=-T_p\int d^{p+1}\xi e^{-\varphi}V(T)\sqrt{-\text{det}\left(\eta_{\mu\nu}+\partial_\mu T\partial_\nu T +\eta_{MN}\partial_\mu X^M\partial_\nu X^N+B_{\mu\nu}+(2\pi\alpha')F_{\mu\nu}\right)},
\end{equation}
where $V(T)$ is the tachyon potential and $\partial_\mu T\partial_\nu T$ is, as usual, the kinetic energy of the tachyons \cite{Garousi1}. Under this conjuncture, the action vanishes at the minimum of the tachyon potential $V(T_0)=0$ \cite{Sent2,Sent1,Recknagel}. %We retrieve Eq.(\ref{tt1a}) at $T=T_0$. 
We will now continue the discussion by approximating that the D$p$-branes are nearly flat, with constant dilaton, and vanishing antisymmetric two-form term $B_{\mu\nu}$ to remove the close string quanta in the system. %\,A^{(p+1)}_{\mu_{1}...\mu_p

\subsection{Dimensional Reduction $\text{U}(1)$ Yang-Mills Theory}\label{DRYMT}
Studying D$p$-branes under Yang-Mills theory enables us to understand the physics of D$p$-branes without necessarily applying any complex string theory artifacts. Detailed analyses show there is enough evidence that super Yang-Mills theory carries a lot of information concerning string theory than one may possibly imagine \cite{Banks}. Besides, recent developments in high-energy physics investigations have shown that string theory gives insight into low-energy field theories in the nonperturbative region \cite{Taylor}. This has been conjectured to be equivalent to QCD where confinement can be realized in {\it color fluxtube} picture. From Eq.(\ref{tt1a}), in the limit of vanishing $B_{\mu\nu}$ and further assuming that the D$p$-branes are almost flat, close to the hypersurface i.e. $X^{M}=0$, $M>p$. Again, we suppose that the fields are slowly varying such that $\partial_\mu X^M\partial^\mu X_M$ and $2\pi\alpha' F_{\mu\nu}$ are in the same order. Therefore, the action can be expanded as
\begin{equation}\label{ymt1}
S=-\tau_pV_p-\dfrac{1}{4g^2_{YM}}\int d^{p+1}\xi\left(F_{\mu\nu}F^{\mu\nu}+\dfrac{2}{(2\pi\alpha')^2}\partial_\mu X^M\partial^\mu X_M \right) +{ {\cal O}((\partial_\mu X^M)^4,F^4)}.
\end{equation}  
Here, $V_p$ is the $p$-brane worldvolume and $g_{YM}$ is the Yang-Mills coupling given by,
\begin{equation}\label{ymt2}
g_{YM}^2=\dfrac{1}{(2\pi\alpha')^2\tau_p}=\dfrac{g}{\sqrt{\alpha'}}\left(2\pi\sqrt{\alpha'}\right)^{p-2}.
\end{equation}
The second term in Eq.(\ref{ymt1}) corresponds to $\text{U}(1)$ gauge theory in $p+1$ dimension coupled with $9-p$ scalar fields. Introducing fermions fields, as mentioned below Eq.(\ref{tt1}) into the action, we recover the supersymmetric $\text{U}(1)$ Yang-Mills theory in $10\,d$ $\mathcal{N}=1$ Super Yang-Mills action,
\begin{equation}\label{ymt3}
S=\dfrac{1}{g_{YM}^2}\int d^{10}\xi\left(-\dfrac{1}{4}F_{\mu\nu}F^{\mu\nu}+\dfrac{i}{2}\bar{\psi}\Gamma^\mu\partial_\mu\psi\right).
\end{equation}
In the case of N parallel D$p$-branes, the p-dimensional branes must be distinctively labeled from $1\,\text{to}\,\text{N}$. Subsequently, the massless scalar fields living on the individual D$p$-brane worldvolume are related to $\text{U}(1)^N$ gauge group. The fields arising from the strings stretching from one brane to the other are labeled as $A^\mu_{i,j}$, where $i,j$ specifies the individual branes that carry the endpoints of the strings. The strings are oriented such that they consist of $N(N-1)$ fields, corresponding to $A_\alpha=(A_\mu,\,X^M)$ individual fields. The mass of the strings is proportional to the separation distances between the branes. So the  strings become massless \cite{Witten1,Taylora} when the D$p$-branes get very close to each other. Hence, the open strings can transform under the adjoint representation $\text{U(N)}$. Thus, the corresponding fields can be decomposed similarly to adjoint supersymmetric gauge field $\text{U(N)}=\text{SU}(N)\times\text{U}(1)$ in $p+1$-dimensions. With the stacks of D$p$-branes crossing each other at some angles, one can also break the $\text{U(N)}$ symmetry into SM particle physics i.e. $\text{SU}(3)\times\text{SU}(2)\times\text{U}(1)$ \cite{Lust}. In sum, the DBI action can be generalized into a non-Abelian gauge group by considering stacks of D$p$-branes instead of a D$p$-brane.

A major motivation for string dual to QCD is derived from the 't Hooft large $N_c$-limit \cite{Hoofta2,Witten4}. Though, $\text{SU}(N_c)$ and $\text{U}(N_c)$ have different representations, in the limit of $N_c\rightarrow\infty$, the difference can be overlooked. Thus, the number of gluons can be approximated as $~N_c^2$. This is more than the quark degree of freedom $N_fN_c$, therefore, we expect the dynamics of the gluons to dominate in this regime \cite{Mateos}. 

%In sum, non-Abelian supersymmetric Yang-Mills theory (SYM) is recovered by $p+1$-dimensional reduction of $10\,d$ non-Abelian YM theory in which the fields are in the adjoint representation of $\text{U(N)}$. Accordingly, after consistent assumptions, the nonperturbative theory describing paralleled D-branes is related to SYM field theory.  

\section{Bosonic Strings at Tachyonic Vacuum}\label{bstv}
{Generally, there are two known boundary conditions associated with open strings. The Dirichlet (fixed) boundary condition is where the coordinates are normal to the brane, and under Neumann (free) boundary condition is where the coordinates are parallel to the brane \cite{Dai,Leigh,Polchinski1}. It has been established \cite{Callan1} that a small disturbance normal to the string and the brane are likely to reflect back with a phase shift of $\pi$, corresponding to a Dirichlet boundary condition. Some study in this regard has been carried out in \cite{Rey,Lee} using Nambu-Goto action for strings with their endpoints on a supergravity background of D$3$-branes. Since the strings attached to the $3$-branes manifest themselves as electric charges, the Neumann boundary condition where the endpoints of the strings are freely oscillating will lead to the production of electromagnetic dipole radiation at the asymptotic outer area of the brane.}

In this section, we will consider Eq.(\ref{tt1aa}) which includes the dynamics of tachyons on open strings. Accordingly, we will keep all the arguments made for the $10$-dimensional DBI action, but we will reduce the spacetime dimension from $10$ to $4$. One of the major differences between superstring theory and particle theory is that the former lives on $10$-dimensional spacetime and the latter on $4$-dimensional spacetime. Nonetheless, this discrepancy can be dealt with using compactification scheme \cite{Gell-Mann,Guendelman,Randall}, where the spacetime is divided into an external non-compact spacetime $\mathcal{M}_{10-d}$ and an internal compact spacetime $\mathcal{M}_d$. We can combine these two phenomena into a single expression as
\begin{equation}
\mathcal{M}_{10}=\mathcal{M}_{10-d}\times\mathcal{M}_d.
\end{equation}
We adopt a physically realistic phenomenon where $d=6$. Additionally, we can set a compact scale $M_c=1/R$, where $R$ is the radius of the internally compact spacetime, smaller than the string mass $M_s=1/l_s$ \cite{Grana, Shiraishi}. As a result, the energy $E$ required for this work must lie in the range $E\ll M_c\ll M_s$. So, the worldvolume coordinate becomes $\xi=(x^\mu,x^6)$, where $\mu=0,1,2,3$.
Consequently, we will reduce the dimension to $1+3$-dimension with $\eta_{\mu\nu}=\text{diag}(+,-,-,-)$ metric signature. We will also decouple the $6$ available transverse fluctuating scalar fields and the antisymmetric tensor i.e., $X^M=B_{\mu\nu}=0$. Keeping the dilaton field constant, we get
%This is in the limit in configuration space where the tachyons decouple from  other massive stringy modes. In that case, the dynamics of the tachyons can be studied by simplifying the action. Consequently, Eq.(\ref{tt1}) can be expressed as,
\begin{align}\label{tt4}
S&=-\tau_p\int d^{4}xV(T)\sqrt{-\text{det}\left(\eta_{\mu\nu}+\partial_\mu T\partial_\nu T +(2\pi\alpha')F_{\mu\nu} \right)}\nonumber\\
&=-\tau_p\int d^{4}xV(T)\sqrt{1-\eta^{\mu\nu}\partial_\mu T\partial_\nu T+\dfrac{1}{2}(2\pi\alpha')^2F_{\mu\nu}F^{\mu\nu}+\cdots}\nonumber\\
&=-\tau_p\int d^{4}xV(T)\left[1-\dfrac{1}{2}\partial_\mu T\partial^\mu T+\dfrac{1}{4}(2\pi\alpha')^2F_{\mu\nu}F^{\mu\nu}+\cdots\right].
\end{align}
%{\color{red} COMMENTS: This is not correct. Indeed it is the third step that is an approximation.} 
{In the above expression we calculate the determinant up to the second-order derivative discarding higher-order corrections}. %This represents a modification of Eq.(\ref{tt1aa}) where the open strings are studied at the tachyonic vacuum. In this region, the strings behave like close strings because the open strings are connected by a flux line on the {D}$p$-brane worldvolume. {\color{blue}As mentioned above, the energy density of the tachyons here, cancels the energy density of the D$p$-branes, $V(T_0)+\tau_p=0$.}{\color{red} COMMENTS: The previous assumption is problematic because it is true for the full theory of string field theory --- see https://arxiv.org/pdf/hep-th/9902105.pdf. The action for the tachyon we are using is an effective action. Please see https://arxiv.org/pdf/hep-th/0303057.pdf and https://arxiv.org/pdf/hep-th/0203265.pdf} %Consequently, the branes dissolve into point-like form. 
{In view of field theory at tachyonic vacuum, the D-branes do not vanish completely rather, there are lower dimensional D-branes present. So, the endpoints of the strings sitting on the low dimensional D-branes behave like point-like particles which serve as the source and sink of the flux carrying the color particles. For instance, expanding around the 'true' minimum of the potential gets rid of the tachyons leading to a particle with a positive square mass.} %Since the negative energy density of the tachyons at the extremum cancels the tension of the branes, there will be no cost of energy in adjusting the strings on the worldvolume of the {D}$p$-brane. %only the tachyonic strings are present. %last step we have set $(2\pi\alpha')=1$.
 
%{\color{red} COMMENTS: This is not correct. Indeed $\phi$ is still the tachyon field in a new variable. This transformation comes from the Lagrangian in $T$ at DBI type action (\ref{tt4}) to $\phi$ at the 'usual' Lagrangian form (\ref{tt7}). Sadly it seems that you never noticed this !} 
{We consider a field configuration such that $T(r)= f(\phi(r))$, in this case, the potential can be expressed as}
\begin{equation}\label{tt5}
V(T(\phi))=\left(\dfrac{\partial\phi}{\partial T} \right)^2,
\end{equation}
so
\begin{align}\label{tt6}
\dfrac{1}{2}V(T)\partial_\mu T\partial^\mu T=\dfrac{1}{2}\left(\dfrac{\partial\phi}{\partial T}\dfrac{\partial T}{\partial x} \right)^2=\dfrac{1}{2} \partial^\mu\phi\partial_\mu\phi.
\end{align}
As a result, the Lagrangian of the system becomes,
\begin{equation}\label{tt7}
\tau^{-1}_p\mathcal{L}=\frac{1}{2}\partial_\mu\phi\partial^\mu\phi-V(\phi)-\frac{1}{4}G(\phi)F_{\mu\nu}F^{\mu\nu},
\end{equation}
here, we have introduced a dimensionless quantity $G(\phi)=(2\pi\alpha')^2V(\phi)$ which will be referred to as a {\it color dielectric function}, subsequently. If we set $(2\pi\alpha')=1$ the string tension becomes $T_{string}=1$ and $G(\phi)=V(\phi)$ \cite{Brito,Adamu,Issifu,Issifu1,Issifu2}. It is important to mention at this point that the potential of the field, $\phi$, follows the same discussion as contained in Sec.~\ref{T}. It has its minimum at $V(\phi=\langle\phi\rangle_0)=0$, where $\langle\phi\rangle_0$ is the 'true' vacuum of the potential. To apply this theory to asymptotically free systems, the potential must satisfy additional conditions,
\begin{equation}\label{tt7a}
V(\phi=\langle\phi\rangle_0)=0\qquad{,}\qquad\dfrac{\partial V}{\partial\phi}|_{\phi=\langle\phi\rangle_0}=0 \qquad{\text{and}}\qquad \dfrac{\partial V}{\partial\phi}|_{\phi=0}=0,
\end{equation}
necessary for stabilizing its vacuum \cite{Rosina,Adamu,Kharzeev}. These restrictions make it possible to apply Eq.(\ref{tt7}) in investigating {\it asymptotically free} particles such as gluons in a QCD-like fashion. Here, the modified Abelian gauge mimics the role of the non-Abelian gauge in the 'usual' QCD theory. We treat the endpoints of the open string on the lower dimensional branes as the source of quark and an anti-quark (if fermions are present) or valence gluons and the string connecting them as gluons that mediate the interactions. Also, the potential $V(\phi)$ plays a similar role as a quantum correction in gluodynamics theory that breaks the conformal symmetry to bring about gluon condensation \cite{Kharzeev, Gaete, Issifu1}. This model has been used to study the color confinement of glueballs at a finite temperature in Ref.\cite{Issifu}.

Considering brane-anti-brane systems, we note that the model must be invariant under the global rotation $\phi\rightarrow e^{i\alpha}\phi$. This warrants an introduction of a complex scalar field $\phi$ with potential $V(|\phi|)$ \cite{Sen3}. %to incorporate their corresponding R-R charge in to the theory. 
So, we can redefine the scalar field in a form, 
\begin{equation}\label{tt8}
\phi=\dfrac{\phi_1+i\phi_2}{\sqrt{2}},
\end{equation}
to incorporate the D$p$-brane and anti-D$p$-brane dynamics. Supposing that the original gauge field $F_{\mu\nu}$ is on the worldvolume of the D$p$-brane  represented by a  complex scalar field $\phi$, we will have a dual gauge field $\tilde{F}_{\mu\nu}$ also on the wordvolume of the anti-D$p$-brane represented by the conjugate field $\phi^*$. Therefore, the string here has its endpoints on remnants of D$p$-brane  and the anti-{D}$p$-brane parallel to each other. %this changes the degrees of freedom of the scalar field appropriately. 
Imposing gauge invariance on the scalar sector, we can define an {\it Abelian covariant derivative}
\begin{equation}\label{tt9}
\partial_\mu\rightarrow D_\mu\equiv \partial_\mu-iq\tilde{A}_\mu,
\end{equation}
where $\tilde{A}_\mu$ is a gauge field which is dual to $A_\mu$. Accordingly, the Lagrangian in Eq.(\ref{tt7}), can be extended as 
\begin{equation}\label{tt10}
\tau_p^{-1}\mathcal{L}=D_\mu\phi D^\mu\phi^*-V(|\phi|)-\frac{1}{4}G(|\phi|)F_{\mu\nu}F^{\mu\nu}-\dfrac{1}{4}\tilde{F}_{\mu\nu}\tilde{F}^{\mu\nu}, 
\end{equation} 
where $F_{\mu\nu}=\partial_\mu A_\nu-\partial_\nu A_\mu$ and $\tilde{F}_{\mu\nu}=\partial_\mu \tilde{A}_\nu-\partial_\nu \tilde{A}_\mu$ are two independent Abelian field strengths. Aside from the original $\text{U}(1)$ gauge invariance of the Lagrangian, it is also invariant under the $\tilde{\text{U}}(1)$ gauge transformation,
\begin{equation}\label{tt11}
\phi(x)\rightarrow\phi'(x)=e^{-iq\alpha(x)}\phi\qquad\text{and}\qquad \tilde{A}_\mu(x)\rightarrow\tilde{A}'_\mu(x)=\tilde{A}_\mu(x)-\partial_\mu\alpha(x).
\end{equation} 
This Lagrangian can undergo a spontaneous symmetry breaking (SSB) process (when we choose an appropriate potential) similar to the usual Abelian Higgs mechanism \cite{Quigg, Collins1}. In this case, $\phi$ plays a role similar to the 'usual' Higgs field in the standard model of particle physics. This process will also lead to the observation of Goldston boson (most likely $\pi^0$ due to the $\tilde{\text{U}}(1)$) corresponding to the number of unbroken symmetries signaling confinement as well \cite{Issifu, Nielsen}. This model has been exploited in detail to study glueballs and color superconductivity in \cite{Issifu2}. { It is important to mention that (\ref{tt10}) is valid when we consider D$p$-brane and an anti-D$p$-brane system in the framework of DBI action \cite{Erkal, Senn}. It is suitable for describing massless particles such as glueballs or gluon confinement.}

\section{Gauge Theories Modified with $G(\phi)$}\label{GTM}
\subsection{Fermions Coupled To Fundamental Strings at Tachyonic Vacuum}\label{mts}
In this section, we introduce Dirac's Lagrangian for free particles adopted by Maxwell in the unification of electric and magnetic field interactions. We introduce the dynamics of the fermions which were dropped in the previous discussions as stated below Sec.~\ref{mdbia}. However, it will be introduced here through Dirac's equation modified by $G(\phi)$ while taking into consideration all {\it gauge invariance} properties.  %One fascinating feature about Maxwell's equations is that, it has no known specific electromagnetic potential that uniquely conform with it to generate fields. Therefore, there is freedom to select different types of potentials to study the same electromagnetic fields, this is generally referred to as {\it gauge invariance}. %We will begin with classical electrodynamic Maxwell's Lagrangian while we modify it with the {\it dielectric function} $G(\phi)$ defined in Sec.~\ref{tts}.
We start with the well-known Dirac's Lagrangian for free particles 
\begin{equation}\label{mt1}
\mathcal{L}_0=\bar{\psi}\left( i\gamma^\mu\partial_\mu-m\right) \psi.
\end{equation}
Even though this Lagrangian is well known in QED, it is also used to describe free nucleons in terms of their composite fermions; protons, and neutrons in strong interaction. It is invariant under global phase rotation,
\begin{equation}\label{mt2}
\psi\rightarrow \psi'= e^{i\alpha}\psi .
\end{equation}
%respectively.
Noticing that the fields in the Lagrangian are color neutral in nature, to apply them in studying color particles such as the ones considered here, we need to modify the fields. Hence, we will modify the bispinors with the $G(\phi)$ in order to give them some color features. Thus, we perform the transformations 
\begin{equation}\label{mt3}
\psi\rightarrow \psi'\equiv G^{1/2}\psi \qquad\text{and}\qquad \bar{\psi}\rightarrow\bar{\psi}'\equiv \bar{\psi}G^{1/2},
\end{equation}
where $G$ is a function %as defined in Eqs.(\ref{tt7}) and (\ref{tt10}) %or complex as in Eq.(\ref{tt10}) depending on the particles under study 
therefore, the gradient in Dirac's equation will transform as 
\begin{equation}\label{mt4}
\partial_\mu\psi\rightarrow\partial_\mu\psi'=\left[ (\partial_\mu G^{1/2})+G^{1/2}\partial_\mu\right] \psi,
\end{equation}
and Lagrangian (\ref{mt1}) becomes 
\begin{align}\label{mt5}
\mathcal{L}&=\bar{\psi}'\left[i\gamma^\mu\partial_\mu-m\right] \psi'\nonumber\\
&=\bar{\psi}\left[ i\gamma^\mu G\partial_\mu+i\gamma^\mu G^{1/2}(\partial_\mu G^{1/2})-mG\right] \psi.
\end{align}
The {\it local gauge invariance} is violated by $G(\phi)$ and the gradient function $(\partial_\mu G^{1/2})$. To ensure local gauge invariance is satisfied, we modify all variables involving derivatives, and color neutral fields such as the Dirac $\gamma$-matrix with $G(\phi)$ and also introduce the electromagnetic field $A_\mu(x)$ in other to make the equation gauge invariant. Consequently, we will adopt the transformations,
\begin{equation}\label{mt6}
\gamma^\mu\rightarrow\gamma'^\mu=G^{-1}\gamma^\mu \qquad{\text{and}}\qquad \partial_\mu\rightarrow D_\mu\equiv \partial_\mu-iqA_\mu,
\end{equation}
where $D_\mu$ is the {\it gauge-covariant derivative} and $q$ is the electric charge. This type of derivative corresponds to momentum transformation, $p_\mu\rightarrow p_\mu-qA_\mu$. By this transformation, the gauge field also gets modified and enters the Lagrangian as $GA_\mu$. It also introduces a coupling between the electromagnetic field and matter in a form $D_\mu\psi$. So Eq.(\ref{mt5}) becomes 
\begin{align}\label{mt7}
\mathcal{L}&=\bar{\psi}\left[ i\gamma^\mu\partial_\mu+i\gamma^\mu\partial_\mu(\ln G^{1/2})+qA_\mu \gamma^\mu-mG\right]\psi\nonumber\\
 &=\bar{\psi}\left[i\gamma^\mu\partial_\mu+q\gamma^\mu A_\mu-mG \right]\psi ,
\end{align}
consequently, the electromagnetic field transforms as 
\begin{equation}\label{mt8}
A_\mu\rightarrow A'_\mu\equiv A_\mu +\dfrac{i}{q}\partial_\mu(\ln G^{1/2}).
\end{equation}
Equation (\ref{mt7}) looks similar to Dirac's Lagrangian with interaction term that couples the gauge field $A_\mu(x)$ to the conserve current $j^\mu=q\bar{\psi}\gamma^\mu\psi$ with a modified mass term $M(r)=mG(\phi)$. %The current $j^\mu$ is conserved under the {\it global symmetry} whilst the {\it local gauge symmetry} leads to the conservation of electronic charge $q$. 
Also, $\bar{\psi}D_\mu\psi$ becomes invariant under local phase rotation, granting interaction between $\bar{\psi}$, $\psi$ and $A_\mu$ with momentum $p_\mu\rightarrow i\partial_\mu$. %The coupling of electromagnetic field and matter is guaranteed by this form of gauge invariance as well. 
In this way, the Lagrangian has been modified, but we ensure that all conservation laws are duly respected.  %Supposing that $G^{1/2}$ is equal to the global phase rotation in Eq.(\ref{tt2}) the 
%gauge transformation becomes 
%\begin{equation}\label{mt9}
%A_\mu\rightarrow A'_\mu\equiv A_\mu -\dfrac{i}{q}\partial_\mu\alpha(x)
%\end{equation}
%this is simply the electromagnetic gauge. The Lagrangian is invariant and under electromagnetic gauge and electromagnetic current is also observed.

Now, to derive a complete Lagrangian that can mimic QCD theory, we include the kinetic energy term of the gauge field which has been derived in Eq.(\ref{tt7}), thus,
%\begin{equation}\label{tt7}
%T_p^{-1}\mathcal{L}=\frac{1}{2}\partial_\mu\phi\partial^\mu\phi-V(\phi)-\frac{1}{4}G(\phi)F_{\mu\nu}F^{\mu\nu}
%\end{equation}
%\begin{equation}\label{mt10}
%\mathcal{L}_{\text{gauge}}=-\dfrac{1}{4}F_{\mu\nu}F^{\mu\nu},
%\end{equation}
%here, the field strength is expressed as $F_{\mu\nu}=\partial_\mu A_\nu-\partial_\nu A_\mu$, putting Eqs.(\ref{mt7}) and (\ref{mt10}) together, we get
\begin{equation}\label{mt11}
\mathcal{L}=\bar{\psi}\left(i\gamma^\mu\partial_\mu+q\gamma^\mu A_\mu-mG(\phi)\right)\psi+\frac{1}{2}\partial_\mu\phi\partial^\mu\phi-V(\phi)-\frac{1}{4}G(\phi)F_{\mu\nu}F^{\mu\nu}.
\end{equation}
This expression looks similar to the usual QED Lagrangian with $\phi$ as intermediate field/particle and a mass term modified by {\it color dielectric function}, $G(\phi)$. As a result, this expression approximates the non-Abelian QCD theory with an Abelian one. This was motivated by the projection that the confining regime of QCD is mostly Abelian dominated \cite{Hoofta1, Ezawa, Shiba, Suzuki, Sakumichi}. The spinor fields $\psi$ and $\bar{\psi}$ represent the quarks and antiquarks and the modified gauge field $A_\mu$ also describes gluons. All the long-distance behavior of the gluons is absorbed in $G(\phi)$. The observed mass of the system $M(r)=mG(\phi)$ %is a function of the tachyonic field $\phi$, 
is expected to have a fixed value at the beginning of the interaction $r\rightarrow r_0$ and at the end of the interaction $r\rightarrow r_*$, so it can be measured precisely. %The behaviour of the particles in the intermediate region is model specific. 
That is, the dielectric function should be such that $G(\phi(r\rightarrow r_0))=G(\phi(r\rightarrow r_*))=\text{constant}$, where $1/r_0$ is the energy at the beginning of the interaction and $1/r_*$ is the energy at the end of the interaction. Thus, $M(r)=mG(\phi)$ is the {\it constituent quark mass function} of the system. %Where, $q$ and $M(r)$ represent the 'physical' electric charge of the electron and its mass with no UV divergences. 
We have presented a detailed study of this theory in Refs.\cite{Issifu1, Adamu} 

Since the {\it renormalization theory} remains the systematic approach for which UV divergences can be resolved \cite{Neubert}, we will compare the result with the renormalized QED Lagrangian to properly identify the nature of the dielectric function in the context of renormalization factor $Z(\mu)$, where $\mu$ is a scale that comes from the dimensional regularization scheme \cite{Bollini, Hoofta}. The objective is to ensure that the result obtained in Eq.(\ref{mt11}) does not pose any UV divergences.  When the {\it real QED} Lagrangian is written in terms of the renormalized factors, it takes the form,
\begin{equation}\label{mt12a}
\mathcal{L}_{\text{QED}}=Z_\psi\bar{\psi}(i\gamma_\mu\partial^\mu-m_0)\psi-\dfrac{Z_A}{4}F^{\mu\nu}F_{\mu\nu}-Z_1 e\bar{\psi}\gamma_\mu A^\mu\psi ,
\end{equation}
where 
\begin{equation}\label{mt13a}
\psi_0=Z_\psi^{1/2}\psi \quad\text{,}\quad A_0^\mu=Z_A^{1/2} A^\mu \quad\text{,}\quad m_0=\dfrac{Z'_\psi}{Z_\psi}m \quad\text{and}\quad e=\dfrac{Z_\psi}{Z_1}Z_A^{1/2}e_0,
\end{equation}
the two equations bear some resemblance, so we can compare them. With $Z_3$ the gluon propagator renormalization factor, $Z_1$ the quark-quark-gluon vertex renormalization, and $Z_2$ the quark self-energy renormalization factor. Additionally, the covariant derivative can be expressed in terms of the renormalized factors as $D^{ren}_\mu\,=\,\partial_\mu-ie(Z_1/Z_\psi)A_\mu$, gauge invariance requires that $Z_1\,=\,Z_\psi$. We have substituted the 'conventional' representation of the renormalization factors  $Z_2$ and $Z_3$ with $Z_\psi$ and $Z_A$ respectively, and $Z_1\,=\,Z_\psi Z_A^{1/2}$ to make the distinction more obvious relative to fermions and the gauge field. The {\it renormalized fields} are without subscript '0' \cite{Itzykson, Pascual, Peskin, Collins, Weinberg, Weinberg1, Schwartz}. Comparing the results in Eqs.(\ref{mt11}) and (\ref{mt12a}), we identify $Z_A\,=\,Z_\psi\,=\,G$ and the gauge invariance warrant that, $Z_A\,=\,1$. Thus, in addition to the color properties carried by the {\it color dielectric function}, it also absorbs the UV divergences. Consequently, the dielectric function follows the restrictions, 
\begin{equation}\label{mm9}
  G\left(\phi(r)\right)\rightarrow 
\begin{cases}
    1 & \text{for}\,\; r\,\rightarrow\,0,\;\text{deconfinement/Coulombian regime}\\
    0 & \text{for} \,  \; r\,\rightarrow\,r_* ,\;\text{confinement regime}
\end{cases}.
\end{equation}
%with observe that the renormalized fields can be related to the fields modified by the dielectric function. 
%Again, $Z_\psi$ can be related to the dielectric function while $Z_AZ_\psi^{1/2}=1$ as can be seen in Eqs.(\ref{mt3}) and (\ref{mt13a}), note that $Z_A$ and $Z_\psi$ are functions of $\mu$. Consequently, the electric charge of the electron $q$ and the mass $M(r)$ in Eq.(\ref{mt11}) can be seen to be 'physical' with no divergences.
From the gauge conditions adopted above, a gluon mass term, 
\begin{equation}\label{mt12}
\mathcal{L}_\gamma=\dfrac{1}{2}M^2(r)A_\mu A^\mu,
\end{equation}
will not be invariant under the local gauge transformation in Eq.(\ref{mt8}) because
\begin{equation}\label{mt13}
A_\mu A^\mu\rightarrow A'_\mu A'^\mu\equiv \left(A_\mu +\dfrac{i}{q}\partial_\mu(\ln G^{1/2}) \right) \left(A^\mu +\dfrac{i}{q}\partial^\mu(\ln G^{1/2}) \right) \neq A_\mu A^\mu.
\end{equation}
Consequently, local gauge invariance accounts for the existence of massless photons \cite{Quigg}.

\subsection{Non-Abelian Gauge Theory}\label{NAGT}
In this section, we will focus on constructing a non-Abelian gauge theory traditionally used for describing the strong nuclear force. Additionally, we buy into the projection that proton and neutron have the same mass and are charge independent due to the strong nuclear force. Therefore, there is an agreement that isospin is conserved in strong interactions. As in the case of QED theory, we will base the discussions on $\text{SU}(2)$-isospin gauge theory introduced by Yang and Mills \cite{Yang}, elaborated by Shaw \cite{Shaw}. Here, we will consider Eq.(\ref{mt1}) as the  Lagrangian for free nucleons with composite fermions, protons (p), and neutrons (n)
\begin{equation}\label{qt1}
\psi\equiv\begin{pmatrix}p \\ n\end{pmatrix}.
\end{equation}
The Lagrangian is invariant under global spin rotation
\begin{equation}\label{qt1a}
\psi\rightarrow\psi'=e^{i\vec{\tau}\cdot\vec{\alpha}/2}\psi,
\end{equation}
also, isospin current, $j^\mu=\bar{\psi}\gamma^\mu({\tau}/{2})\psi$, is conserved therefore, proton and neutron can be treated symmetrically in the absence of electromagnetic interactions. In that case, the distinction between proton and neutron is arbitrary and conventional under this representation. 

To maintain the differences between the Abelian theory treated in Sec.~\ref{mts} and the non-Abelian theory intended for this section, we will replace $A_\mu$ from the {\it Abelian covariant derivative}, $D_\mu$, expressed in Eq.(\ref{mt6}) with its non-Abelian counterpart $B_\mu$ i.e.,
\begin{equation}\label{qt1b}
D_\mu\equiv \partial_\mu-igB_\mu.
\end{equation}
We have also replaced $q$ with $g$, the strong coupling constant to make the analysis distinct from Sec.~\ref{mts} and befitting for describing strong interactions. Despite the similar global gauge invariance satisfied by both theories, they also exhibit some major differences; the algebra of the non-Abelian group is more complex and the associated gauge bosons self-interact due to the structure of the non-Abelian group. Accordingly, the field can be expressed as,
\begin{equation}\label{qt2}
B_\mu=\dfrac{1}{2}\vec{\tau}\cdot\vec{b}_\mu=\dfrac{1}{2}\tau^ab^a_\mu=\dfrac{1}{2}\left( {\begin{array}{cc}
   b_\mu^3 & b^1_\mu-ib^2_\mu \\
   b^1_\mu +ib^2_\mu& -b^3_\mu \\
  \end{array} } \right)=\dfrac{1}{2}\left( {\begin{array}{cc}
   b_\mu^3 & \sqrt{2}\,b^+_\mu \\
   \sqrt{2}\,b^-_\mu & -b^3_\mu \\
  \end{array} } \right),
\end{equation}
where the three non-Abelian gauge fields are $\vec{b}_\mu=(b_\mu^1,b_\mu^2,b_\mu^3)$. The generators $\tau^a$ $(a=1,\,2$ and $3$) are associated with Pauli's matrices, $b_\mu^{\pm}=(b_\mu^1\mp ib_\mu^2)/\sqrt{2}$ are the charged gauge bosons and the isospin step-up and step-down operators $1/2(\tau_1\pm i\tau_2)$ exchange $p\leftrightarrow n$ following the absorption of $b_\mu^{\pm}$ boson. The gradient in Eq.(\ref{mt1}) will then transform as,
\begin{align}\label{qt3}
D_\mu\psi\rightarrow D\psi'&=G^{1/2}\partial_\mu\psi+(\partial_\mu G^{1/2})\psi-igB_\mu(G^{1/2}\psi)\nonumber\\
&=G^{1/2}\left(\partial_\mu-igB'_\mu\right) \psi,
\end{align} 
thus,
\begin{align}\label{qt3a}
-igG^{1/2}B'_\mu&=-igB_\mu G^{1/2}+(\partial_\mu G^{1/2})\rightarrow\nonumber\\
B'_\mu&=G^{1/2}B_\mu G^{-1/2}+\dfrac{i}{g}(\partial_\mu G^{1/2})G^{-1/2}\nonumber\\
&=G^{1/2}\left(B_\mu+\dfrac{i}{g}G^{-1/2}(\partial_\mu G^{1/2}) \right) G^{-1/2}.
\end{align} 
Again, following similar procedure as adopted for Eq.(\ref{mt7}) using the necessary transformations, we obtain
\begin{align}\label{qt4}
\mathcal{L}'=\bar{\psi}\left[i\gamma^\mu\partial_\mu+gB_\mu\gamma^\mu-mG \right]\psi,
\end{align}
with gauge invariant transformation similar to Eq.(\ref{mt8}), 
\begin{equation}\label{qt5}
B_\mu\rightarrow B'_\mu\equiv B_\mu+\dfrac{i}{g}(\partial_\mu\ln G^{1/2}).
\end{equation}
By comparing Eq.(\ref{qt3a}) and Eq.(\ref{qt5}) we can deduce,
\begin{equation}\label{qt5a}
B_\mu\rightarrow B'_\mu=G^{1/2}B_\mu G^{-1/2}.
\end{equation}
The $\text{SU}(2)$-isospin generators can be expressed in terms of commutator relation, 
\begin{equation}\label{qt7}
\left[\tau^j,\tau^k \right]=2i\varepsilon_{jkl}\tau^l. 
\end{equation}
Conventionally, $\tau_i$ do not commute with others in different spatial directions, so only one component can be measured at a time and this is taken to be the third component $T_3=1/2\tau_3$, generally in $z$-direction. 

We will now develop the field strength tensor that will form the kinetic term of the gauge field. Starting with the electromagnetism gauge, we can construct, 
\begin{equation}\label{qt8}
F_{\mu\nu}=\dfrac{1}{2}{\bf F}_{\mu\nu}\cdot{\bf \tau}=\dfrac{1}{2}F^a_{\mu\nu}\tau^a \qquad{\text{where}}\qquad tr\left( \tau^a\tau^b\right)=2\delta^{ab}.
\end{equation}  
From these transformations, we can develop a gauge invariant field, 
\begin{equation}\label{qt9}
\mathcal{L}'_{gauge}=-\dfrac{1}{4}{\bf F}_{\mu\nu}\cdot{\bf F}^{\mu\nu}=-\dfrac{1}{2}tr\left( F_{\mu\nu}F^{\mu\nu}\right),
\end{equation}
where the field strength tensor transforms under the dielectric function as
\begin{equation}\label{qt10}
F_{\mu\nu}\rightarrow F'_{\mu\nu}\equiv G^{1/2}F_{\mu\nu}G^{-1/2}.
\end{equation}
%here $G$ is a function. %For clarity one can deduce directly from Eq.(\ref{qt3}) that Eq.(\ref{qt5}) can be rewritten in more general form as
%\begin{equation}\label{qt11a}
%B'=G^{1/2}BG^{-1/2}+\dfrac{i}{g}(\partial_\mu G^{1/2})G^{-1/2}
%\end{equation}
%which is the expression that will be used in subsequent computations. 
We know from the QED relation constructed in Sec.~\ref{mts} that the field strength can be expressed as
\begin{align}\label{qt11}
F'_{\mu\nu}&=\partial_\nu B'_\mu-\partial_\mu B'_\nu=\partial_\nu\left[G^{1/2}B_\mu G^{-1/2}+\dfrac{i}{g}(\partial_\mu G^{1/2})G^{-1/2}\right]-\partial_\mu\left[G^{1/2}B_\nu G^{-1/2}+\dfrac{i}{g}(\partial_\nu G^{1/2})G^{-1/2} \right] \nonumber\\
&=\left\lbrace (\partial_\nu G^{1/2})B_\mu G^{-1/2}+G^{1/2}(\partial_\nu B_\mu)G^{-1/2}+G^{1/2}B_\mu(\partial_\nu G^{-1/2})+\dfrac{i}{g}\left[\partial_\nu(\partial_\mu G^{1/2})G^{-1/2}+(\partial_\mu G^{1/2})(\partial_\nu G^{-1/2}) \right]  \right\rbrace \nonumber\\&- \left\lbrace (\partial_\mu G^{1/2})B_\nu G^{-1/2}+G^{1/2}(\partial_\mu B_\nu)G^{-1/2}+G^{1/2}B_\nu(\partial_\mu G^{-1/2})+\dfrac{i}{g}\left[\partial_\mu(\partial_\nu G^{1/2})G^{-1/2}+(\partial_\nu G^{1/2})(\partial_\mu G^{-1/2}) \right]  \right\rbrace .
\end{align}
We adopted the expression for $B'_\mu$ defined in Eq.(\ref{qt3a}), regrouping the terms in the above equation yields,
\begin{align}\label{qt12}
F'_{\mu\nu}&=G^{1/2}\left[\partial_\nu B_\mu-\partial_\mu B_\nu \right]G^{-1/2}+\left[(\partial_\nu G^{1/2})B_\mu-(\partial_\mu G^{1/2})B_\nu \right]G^{-1/2}+ G^{1/2}\left[B_\mu(\partial_\nu G^{-1/2})-B_\nu(\partial_\mu G^{-1/2}) \right]\nonumber\\ &+\dfrac{i}{g}\left[(\partial_\mu G^{1/2})(\partial_\nu G^{-1/2})- (\partial_\nu G^{1/2})(\partial_\mu G^{-1/2})\right] \nonumber\\
&\neq G^{1/2}\left[\partial_\nu B_\mu-\partial_\mu B_\nu \right]G^{-1/2},
\end{align}
higher derivative terms were discarded. We can cast this result in a slightly symmetric form by using the identity $G^{1/2}G^{-1/2}=\mathbb{I}$, so
\begin{align}\label{qt13}
\partial_\mu(G^{1/2}G^{-1/2})&=(\partial_\mu G^{1/2})G^{-1/2}+G^{1/2}(\partial_\mu G^{-1/2})=0 \rightarrow\nonumber\\
G^{1/2}(\partial_\mu G^{-1/2})&=-(\partial_\mu G^{1/2})G^{-1/2}.
\end{align}
Using this identity appropriately leads to, 
\begin{align}\label{qt14}
F'_{\mu\nu}&=G^{1/2}\left(\partial_\nu B_\mu-\partial_\mu B_\nu \right) G^{-1/2}+\left((\partial_\nu G^{1/2})B_\mu-B_\mu(\partial_\nu G^{1/2}) \right) G^{-1/2}+\left(B_\nu (\partial_\mu G^{1/2})- (\partial_\mu G^{1/2})B_\nu\right) G^{-1/2}\nonumber\\
 &+\dfrac{i}{g}\left((\partial_\nu G^{1/2})G^{-1/2}(\partial_\mu G^{1/2})-(\partial_\mu G^{1/2})G^{-1/2}(\partial_\nu G^{1/2})\right) G^{-1/2} \nonumber\\
 %&=G^{1/2}\left(\partial_\nu B_\mu-\partial_\mu B_\nu \right) G^{-1/2}+\left[(\partial_\nu G^{1/2}),B_\mu \right] G^{-1/2}+\left[ B_\nu,(\partial_\mu G^{1/2}) \right] G^{-1/2}\nonumber\\ &+ \dfrac{i}{g}\left[(\partial_\nu G^{1/2}),G^{-1/2}(\partial_\mu G^{1/2}) \right] G^{-1/2}\nonumber\\
 &=G^{1/2}\left(\partial_\nu B_\mu-\partial_\mu B_\nu \right)G^{-1/2}+G^{1/2}\left\lbrace \left[G^{-1/2}(\partial_\nu G^{1/2}),B_\mu \right]-\left[G^{-1/2}(\partial_\mu G^{1/2}),B_\nu \right]\right\rbrace G^{-1/2}\nonumber\\ &+ \dfrac{i}{g}G^{1/2}\left[G^{-1/2}(\partial_\nu G^{1/2}),G^{-1/2}(\partial_\mu G^{1/2}) \right] G^{-1/2}.
\end{align}
%we have made use of the identity $G^{1/2}G^{-1/2}=\mathbb{I}$ in the last step. 
This equation shows the additional terms that come from the non-vanishing commutators due to the non-Abelian group structure. By this expression, we deduce that a term can be added to $\partial_\nu B_\mu-\partial_\mu B_\nu $ to modify $F_{\mu\nu}$ to achieve the desired transformation property we require. With this inspiration, the observed electromagnetic field strength tensor can be modified to read,
\begin{equation}\label{qt15}
F_{\mu\nu}=\dfrac{1}{iq}\left[D_\nu , D_\mu \right]. 
\end{equation}
Substituting the definition for $D_\mu$ in Eq.(\ref{mt6}) into the above expression, we obtain
\begin{equation}\label{qt16}
F_{\mu\nu}=\partial_\mu A_\nu-\partial_\nu A_\mu+iq\left[ A_\nu ,A_\mu \right].
\end{equation}
The commutator vanishes for Abelian theories. Applying this transformation to Eq.(\ref{qt10}) yields,
\begin{equation}\label{qt17}
F'_{\mu\nu}=\partial_\nu B'_\mu -\partial_\mu B'_\nu+ig\left[B'_\nu ,B'_\mu \right].
\end{equation}
Expanding the commutator using the transformation Eq.(\ref{qt3a}) to enable us to compare the outcome to the nonvanishing commutator relations in Eq.(\ref{qt14}) leads to,
\begin{align}\label{qt18}
ig\left[ B'_\nu , B'_\mu \right]&=ig\left[ \left( G^{1/2}B_\nu G^{-1/2}+\dfrac{i}{g}(\partial_\nu G^{1/2})G^{-1/2}\right),\left( G^{1/2} B_\mu G^{-1/2}+\dfrac{i}{g}(\partial_\mu G^{1/2})G^{-1/2}\right)  \right] \nonumber\\ 
%&=ig G^{1/2}\left[B_\nu ,B_\mu \right]G^{-1/2}-G^{1/2}\left[B_\nu ,G^{-1/2}(\partial_\mu G^{1/2}) \right] G^{-1/2}-G^{1/2}\left[G^{-1/2}(\partial_\nu G^{1/2}), B_\mu \right]G^{-1/2} \nonumber\\&- \dfrac{i}{g}G^{1/2}\left[ G^{-1/2}(\partial_\nu G^{1/2}),G^{-1/2}(\partial_\mu G^{1/2})\right] G^{-1/2} \nonumber\\
&=igG^{1/2}\left[ B_\nu , B_\mu \right]G^{-1/2}-G^{1/2}\left\lbrace \left[ G^{-1/2}(\partial_\nu G^{1/2}),B_\mu \right]-\left[ G^{-1/2}(\partial_\mu G^{1/2}),B_\nu \right]\right\rbrace  G^{-1/2}\nonumber\\&-\dfrac{i}{g}G^{1/2}\left[G^{-1/2}(\partial_\nu G^{1/2}),G^{-1/2}(\partial_\mu G^{/2}) \right] G^{-1/2}.
\end{align}
The commutator relations that come after the first term in the last step are the exact terms required to cancel the extra terms in Eq.(\ref{qt14}). Therefore, the field strength tensor expressed in Eq.(\ref{qt17}) has the required structure under local gauge transformation. Combining Eqs.(\ref{qt4}) and (\ref{qt9}) gives rise to modified Yang-Mills Lagrangian \cite{Quigg,Yang},
\begin{equation}\label{qt19}
\mathcal{L}_{\text{YM}}=\bar{\psi}\left(i\gamma^\mu\partial_\mu+g\gamma^\mu B_\mu-mG \right)\psi-\dfrac{1}{2}tr\left(F_{\mu\nu}F^{\mu\nu} \right),
\end{equation}
where $M(r)=mG(\phi)$ is the modified nucleon mass. This Lagrangian is also invariant under local gauge transformations and does not permit the existence of mass term $M^2(r)B_\mu B^\mu$. It should also be noted that introducing the non-Abelian gauge leads to the automatic cancellation of the color dielectric function. Hence, we can infer that the color dielectric function attached to the Abelian gauge induces strong interaction properties. Using the expression in Eq.(\ref{qt2}) and the commutator relation in Eq.(\ref{qt7}), we can rewrite the transformed version of Eq.(\ref{qt17}) as 
\begin{align}\label{qt20}
F_{\mu\nu}&=\partial_\mu B_\nu-\partial_\nu B_\mu +ig\left[ B_\mu ,B_\nu\right]\rightarrow\nonumber\\
F^l_{\mu\nu}&= \dfrac{1}{2}\left(\partial_\mu(\tau^l b^l_\nu)-\partial_\nu (\tau^l b^l_\mu)\right) +\dfrac{ig}{4}\left(-2i\varepsilon_{jkl}(b^j_\nu b^k_\mu \tau^l) \right) \nonumber\\
&=\partial_\mu b^l_\nu-\partial_\nu b^l_\mu+g\varepsilon_{jkl} b^j_\nu b^k_\mu,
\end{align}
in the last step we have dropped the three isospin generators $\tau^l$ of the gauge field because they are linearly independent. Generally, non-Abelian gauge groups that do not fall under $\text{SU}(2)$, the Levi-Civit\'a symbol $\varepsilon_{jkl}$ is replaced with the {\it antisymmetric structure constant} $f_{jkl}$.

The introduction of the $\text{SU}(2)$-isospin symmetry by Heisenberg \cite{Heisenberg} preceded the development of the quark model. That notwithstanding, the only known fundamental components of the nucleon describing strong interactions are up-quark ($u$) and down-quark ($d$). While the proton is composed of two $u$-quarks and a $d$-quark, the neutron is also composed of two $d$-quarks and an $u$-quark. Indeed, these particles remain the only known constituents of proton and neutron with almost the same mass and coupling force. Hence, Eq.(\ref{qt19}) can be used to study the behavior of quarks and gluons inside hadrons. In that case, the up/down quarks are treated  as having the same mass and $0$-charge, so they are seen as similar particles with different isospin states, $I_{3 u/d}=\pm 1/2$, same as the nucleon. Interestingly, the spin addition of the quark constituents of proton and neutron agree with $I_{3}=1/2$ and $I_{3}=-1/2$ respectively. Similar to the isospin representation of the nucleon field $\psi$ in Eq.(\ref{qt1}) the quark field can be represented as
\begin{equation}\label{qt21}
\psi_q\equiv\begin{pmatrix}u \\ d\end{pmatrix}.
\end{equation}
Nevertheless, the mathematical structure of this theory is the same as QCD, the theory that describes the characteristics of quarks and gluons inside the hadrons. 

{Again, the strong interaction is well known for its invariance under quark color permutations, so we can express the quark wave function for color triplet representation as
\begin{equation}\label{qt21a1}
\psi_q\equiv\begin{pmatrix}R \\ G\\B\end{pmatrix},
\end{equation}
where $\text{R}\equiv\text{red}$, $\text{G}\equiv\text{green}$ and $\text{B}\equiv\text{blue}$. This is invariant under $\text{SU}(3)$ transformation of the form
\begin{equation}\label{qt22}
\psi\rightarrow\psi'\equiv \exp\left(i\dfrac{1}{2}\lambda_j\alpha_j \right)\psi ,
\end{equation}
where $\alpha_j, \, j=1,\,2,\,3,\cdots,8$ are the eight phase angles and $\lambda_j$ is a $3\times 3$ matrix representing eight independent traceless Hermitian generators of the group. Ignoring the differences in quark masses, it can also be applied in studying u, d and s quark systems,
\begin{equation}\label{qt21aa}
\psi_q\equiv\begin{pmatrix}u \\ d\\s\end{pmatrix}.
\end{equation}
The generators are fundamentally equivalent to Pauli's matrices for the $\text{SU}(2)$ representation and they satisfy the Lie algebra, 
 \begin{equation}\label{qt23}
 \left[\lambda_i,\lambda_j \right]=2if_{ijk}\lambda_k,
 \end{equation}
similar to Eq.(\ref{qt7}), $f_{ijk}$ is the {\it structure constant}. Consequently, to generalize the relation in Eq.(\ref{qt20}) for QCD under the $\text{SU}(3)$ color symmetry, we substitute the antisymmetric tensor $\varepsilon_{ijk}$ for the {\it antisymmetric structure constant} $f_{ijk}$ \cite{Collins1}.

}
Finally, the models exhibit the expected asymptotic free \cite{Gross, Politzer} behavior at high energy regions while at low energies {\it color confinement} and hadronization sets in.  %making classical description of quantum fields unsuitable. 
Here, $M(r)$ represents {\it constituent quark mass} function \cite{Atkinson} while $m$ is the {\it bare quark mass}. As analyzed in Sec.~\ref{mts}, the constituent mass of the quarks in the asymptotically free region can be determined as $M(r\rightarrow r_0)=M_0$, while the constituent mass at the nonperturbative (low energy) region where {\it color confinement} is expected, will be $M(r\rightarrow r_*)=M_*$. Hence, it is possible to have the same {\it constituent quark mass} in both the UV ($r\rightarrow r_0$) and the IR ($r\rightarrow r_*$) regimes, if indeed the {\it bare quark mass} $m$ in both regimes are the same \cite{Adamu,Eichten}, because the $G(\phi)$ behaves as, $G(\phi(r\rightarrow r_0))=G(\phi(r\rightarrow r_*))=\text{constant}$. However, there is evidence that $m$ is small in the IR regime and large in the UV regime \cite{Adamu, Issifu1}. Accordingly, if we require {\it color confinement} in both regions, $M_0\,>\,M_*$ because higher {\it bare quark mass} $m$ is required to obtain confinement in the UV region than it is required in the IR region.
 
 \section{Phenomenology of Glueball Confinement}\label{TM}
In this section we will use one of the models built in Sec.~\ref{bstv} specifically, Eq.(\ref{tt7}) to build a model that describes the features of confining glueballs. Apart from the insight into the behavior of the glueballs in both the IR and the UV regions, it will serve as a test to the models developed earlier. The model will be based on electric field confinement, commonly referred to as the {\it chromoelectric flux} confinement. In that light, we will define the indices of the gauge field such that the {\it chromomagnetic flux} is eliminated from the system (i.e. $j^\nu=(\rho,\,\vec{0})$) and only the static sector of the scalar field, i.e. $\mu=j$, is available for analysis. This ensures that the color particles that generate the gluons are static. This section will enable us to see how the Abelian gauge can be used to approximate a non-Abelian theory. Additionally, the color dielectric function $G(\phi)$ absorbs the long-distance dynamics of the gluons such that the photon propagator emanating from the Abelian gauge field does not decouple at longer wavelengths. We will further demonstrate that the $G(\phi)$ is directly related to the QCD $\beta$-function and the strong running coupling constantly.

\subsection{The Model}\label{TMA}
The equations of motion for Eq.(\ref{tt7}) are,
\begin{equation}\label{cgt2}
\partial_\mu\partial^\mu\phi+\dfrac{1}{4}\dfrac{\partial G(\phi)}{\partial\phi}F_{\mu\nu}F^{\mu\nu}+\dfrac{\partial V}{\partial\phi}=0,
\end{equation}
and 
\begin{equation}\label{cgt3}
\partial_\mu[G(\phi)F^{\mu\nu}]=0.
\end{equation}
Expressing the above equations in spherical coordinates,
\begin{equation}\label{cgt4}
\dfrac{1}{r^2}\dfrac{d}{d r}\left(r^2\dfrac{d\phi}{dr}\right) =-\dfrac{1}{2}\dfrac{\partial G(\phi)}{\partial\phi}E^2+\dfrac{\partial V(\phi)}{\partial\phi},
\end{equation}
where, $F^{\mu\nu}F_{\mu\nu}=F^{0j}F_{0j}+F^{i0}F_{i0}=-2E^2$ (only electric field components are considered) and $V=G$ as discussed in Sec.~\ref{bstv}. %We define the indices such that $\nu=0$ and $\mu=j=1,\,2\,\text{and}\,3$. 
Accordingly,
\begin{align}\label{cgt5}
\dfrac{1}{r^2}\dfrac{d}{d r}\left(r^2\dfrac{d\phi}{dr}\right)&=\dfrac{\partial}{\partial\phi}\left[ \dfrac{\lambda^2}{2}\dfrac{1}{V(\phi)}\dfrac{1}{r^4}+V \right]\rightarrow\nonumber\\ 
\nabla^2\phi&=\dfrac{\partial}{\partial\phi}\left[ \dfrac{\lambda^2}{2}\dfrac{1}{V(\phi)}\dfrac{1}{r^4}+V \right],
\end{align}
where 
\begin{equation}\label{cgt6}
E=\dfrac{\lambda}{r^2 G(\phi)},
\end{equation}
and 
\begin{equation}\label{cgt7}
\lambda=\dfrac{q}{4\pi},
\end{equation}
$\lambda$ is an integration constant, we also substituted $\varepsilon_0=1$ to achieve the desired objective. Now we choose a potential that satisfies all the conditions expressed in Sec.~\ref{bstv} thus,
\begin{equation}\label{cgt8}
V(\phi)=\dfrac{\rho}{4}[(\alpha\phi)^2-a^2]^2,
\end{equation}
where $\rho$ and $\alpha=1/f_\alpha$ are dimensionless constants, $f_\alpha$ is the tachyon decay constant. This potential contains tachyonic mode at $V''(0)$, so the fields cannot be quantized around this point ---  see Sec.~\ref{T} for detailed discussions. However, we can remove the tachyonic modes by shifting the vacuum $\phi\rightarrow \phi_0+\eta$ and quantizing around the true minimum, $\phi_0=\pm a/\alpha$. Consequently,
\begin{equation}\label{p1}
m_\phi^2=\dfrac{\partial^2 V}{\partial\phi^2}\bigg\vert_{\phi=\phi_0}=2\rho\alpha^2a^2,
\end{equation} 
and the potential can be expanded for the small perturbation $\eta$, which will be referred to as a glueball field with a real square mass, $m_\phi^2$, hence,
\begin{align}\label{p2}
V(\eta)=G(\eta)&=V(\phi)\bigg\vert_{\phi_0}+\dfrac{\partial V}{\partial\phi}\bigg\vert_{\phi_0}\eta+\dfrac{1}{2}\dfrac{\partial^2 V}{\partial\phi^2}\bigg\vert_{\phi_0}\eta^2\nonumber\\
&=\dfrac{1}{2}m_\phi^2\eta^2.
\end{align}
Considering that the particles are sufficiently separated such that {\it color confinement} can be observed, we ignore the $1/r^4$ term in Eq.(\ref{cgt5}) and simplify it as
\begin{align}\label{cgt10}
\nabla^2(\eta)&=\dfrac{\partial V}{\partial\phi}\bigg\vert_{\phi_0}+\dfrac{\partial^2 V}{\partial\phi^2}\bigg\vert_{\phi_0}\eta\rightarrow\nonumber\\
\nabla^2\eta&=\dfrac{\partial^2 V}{\partial\phi^2}\bigg\vert_{\phi_0}\eta,
\end{align}
leading to
\begin{equation}\label{cgt11}
\eta''(r)+\dfrac{2}{r}\eta'(r)-m^2_\phi\eta=0.
\end{equation}
This equation has several solutions but we choose two of such solutions suitable for the analysis,
\begin{equation}\label{cgt12}
\eta(r)=\dfrac{a\cosh(m_\phi r)}{\alpha m_\phi r} \qquad\text{and}\qquad \eta(r)=\dfrac{a\sin(m_\phi r)}{\alpha m_\phi r}.
\end{equation}
Each solution corresponds to the characteristics of the particle in a particular regime i.e., IR and UV regimes respectively. The solution at the IR regime will give rise to a linear confining potential and the UV solution will lead to a Cornell-like potential. 

\subsection{Confining Potentials}\label{CP}
In this section we will present the confining potentials derived from the model by considering the electrodynamic potential,
\begin{equation}\label{p3}
V=\mp\int Edr.
\end{equation}
Substituting the equation at the left side of the solution Eq.(\ref{cgt12}) and Eq.(\ref{cgt6}) into Eq.(\ref{p3}) leads to
\begin{align}\label{p4}
V_c(r)&=\dfrac{2\lambda\alpha^2\tanh(m_\phi r)}{a^2 m_\phi}+c\nonumber\\
&=m_\phi\lambda\tanh(m_\phi r)\qquad\text{for}\qquad a^4\rho=1
\end{align}
where $c$ is an integration constant that is set to zero in the last step. Considering that $m_\phi r\ll 1$, $c=0$ and $\lambda=1$ corresponding to the positive part of the potential $V$, we can deduce the QCD string tension $\sigma$ to be, 
\begin{equation}\label{p5}
\sigma_L={m_\phi^2}.
\end{equation} 
Here, we can infer that confinement is occasioned by the magnitude of the glueball mass, in the limit of vanishing glueball mass there will be no confinement. Furthermore, this potential leads to linear growth in $r$, and at some critical distance, $r_*=1/\sqrt{\sigma_L}$ \cite{Bali} the potential begins to flatten up leading to hadronization. It is known from the flux tube models for confining color particles that, $r\gg r_*$ \cite{Bali}.
\begin{figure}[H]
  \centerline{\includegraphics[scale=0.7]{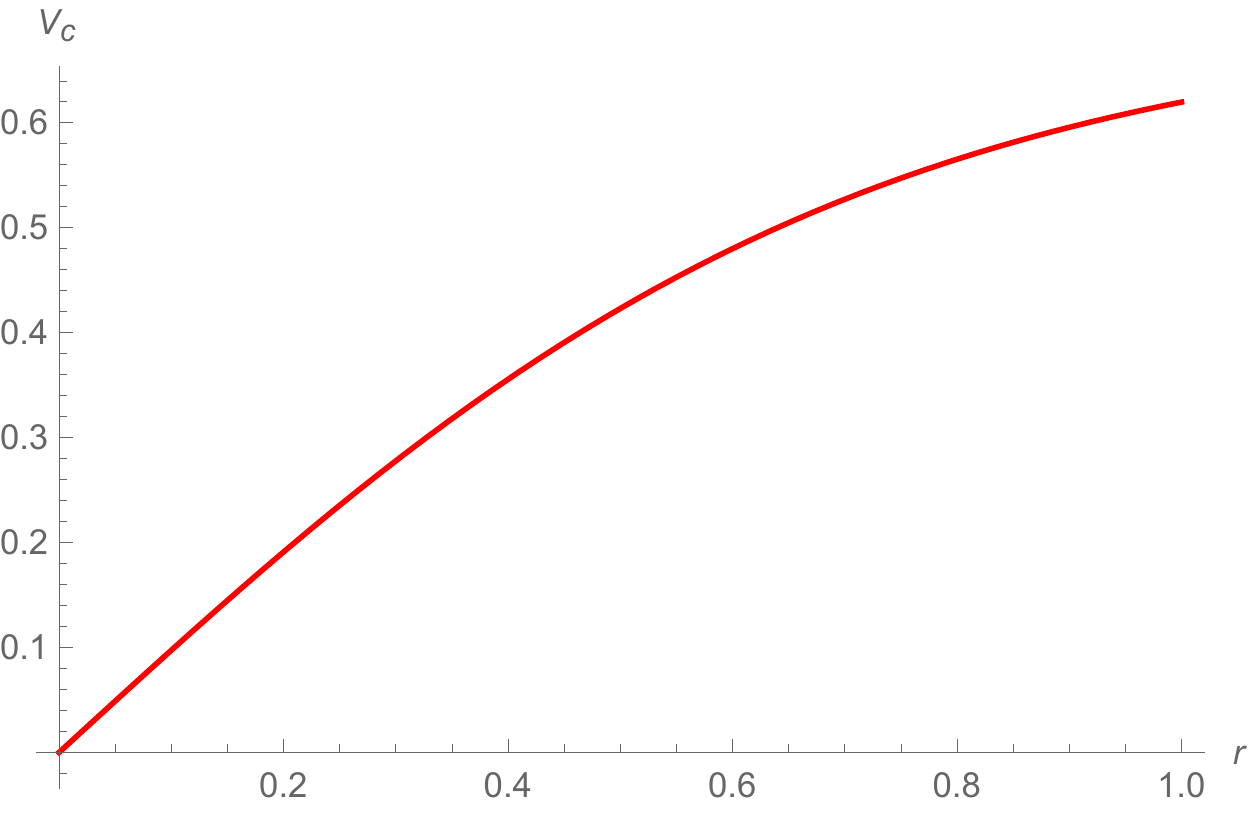}}
  \caption{Flux tube-like potential}
   \label{pa1}
    \floatfoot{Increase in distance $r$ increases the strength of confinement until $r\geq r_*$ where the curve is expected to start flattening, signaling hadronization.}
\end{figure}

On the other hand, taking the solution at the right side of Eq.(\ref{cgt12}) and following the same process as followed above, we get 
\begin{align}\label{cgt16}
V_s(r)&=-\dfrac{2\lambda\alpha^2\cot(m_\phi r)}{a^2m_\phi}+c\nonumber\\
&\simeq -\dfrac{2\lambda\alpha^2}{a^2m_\phi}\left[\dfrac{1}{m_\phi r}-\dfrac{m_\phi r}{3}-{\cal O}(r^3) \right]+c \nonumber\\
&\simeq \dfrac{2\lambda\rho \alpha^2a^2}{\rho a^4m^2_\phi}\left[ -\dfrac{1}{r}+\dfrac{m_\phi^2 r}{3}\right],
\end{align}
in the last step, we set the integration constant $c=0$, also choosing the positive part of the potential corresponding to $\lambda=1$ and $\rho a^4=1$, we arrive at the Cornell-like potential for confining heavy quarks i.e.,
\begin{equation}
V_s(r)=-\dfrac{1}{r}+\dfrac{m^2_\phi r}{3},
\end{equation}
corresponding to a string tension 
\begin{equation}\label{p6}
\sigma_s=\dfrac{m_\phi^2}{3}.
\end{equation}
It is important to recognize that the critical distance, $r_{*s}=1/\sqrt{\sigma_s}$, in this regime marks the transition from the asymptotic freedom region to the confining region.
\begin{figure}[H]
  \centerline{\includegraphics[scale=0.7]{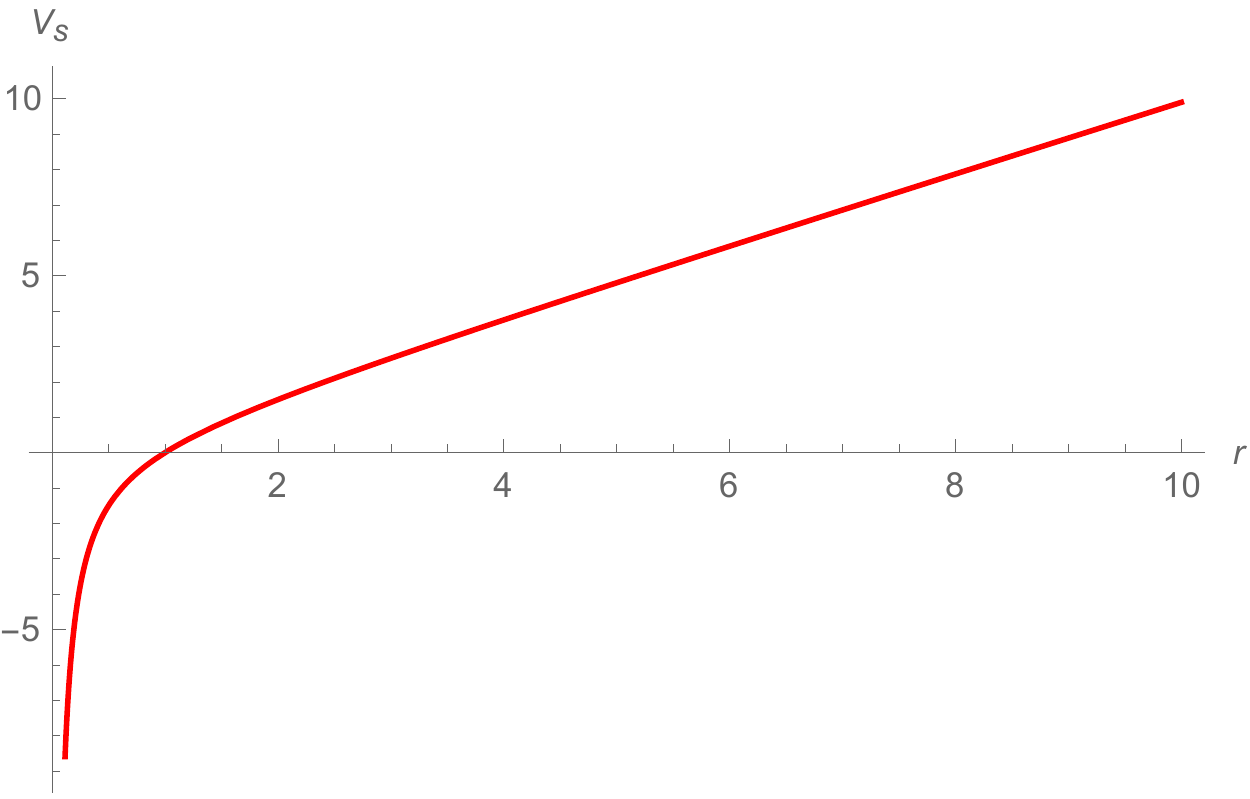}}
  \caption{A graph of Cornell-like potential}
   \label{pa2}
    \floatfoot{An increase in the distance also leads to an increase in the strength of confinement.}
\end{figure}
%More details on this subject can be found in Refs.\cite{Issifu,Adamu,Issifu1,Issifu2}. 
We know from Sec.~\ref{mdbia} that the string tension $T_{\text{string}}\sim\sigma_L\sim\sigma_s\sim 1\, \text{GeV}/\text{fm}$, hence the glueball mass in the IR regime becomes $m_L\approx 1\,\text{GeV}$ corresponding to glueball mass of isoscalar resonance $f_0(980)$ \cite{Tanabashi}. The commonly known ratio of $m(0^{++})/\sqrt{\sigma_L}$ in QCD theory in $\text{SU}(\infty)$ limit \cite{ Albanese,Bacilieri,Teper}, at this regime, can be determined as $m_L/\sqrt{\sigma_L}\approx 1$. Likewise in the UV regime, we have a glueball of mass $m_s\approx1.73\,\text{GeV}$ corresponding to the lightest scalar glueball mass of resonance $f_0(1710)$. The result obtained here is precisely the same as the results obtained from QCD lattice calculations \cite{Tanabashi, Morningstar, Loan, Chen, Lee1, Bali1} also, $m_s/\sqrt{\sigma_s}\approx1.73$. The critical distances become $r_{*s}=r_*=1\,\text{fm}$ for both the IR and the UV regimes. While critical distance in the IR regime $r_*$ refers to the transition from confinement to hadronization regions, critical distance in the UV regime $r_{*s}$ refers to the transition from the asymptotically free region to the confining region. 

\subsection{Gluon Condensation}\label{GC}
Classical theory for gluodynamics is invariant under the scale transformation $x\rightarrow\lambda x$, this leads to a scale current $s_\mu$ which is related to the energy momentum tensor trace $\theta^\mu_\mu(x)$ as 
\begin{equation}
\partial^\mu s_\mu=\theta^\mu_\mu.
\end{equation}
In the absence of quantum corrections, $\theta^\mu_\mu=0$, the theory remains conformally invariant. This will lead to a vanishing gluon condensation $\langle F_{\mu\nu}^aF^{a\mu\nu}\rangle=0$. On the other hand, when quantum correction, $-|\varepsilon_v|$, is introduced, the conformal symmetry is broken leading to non-vanishing gluon condensate $\langle F_{\mu\nu}^aF^{a\mu\nu}\rangle\neq 0$ and energy-momentum trace anomaly comes to play
\begin{equation}\label{p7a}
\theta^\mu_\mu=\dfrac{\beta(g)}{2g}F^a_{\mu\nu}F^{a\mu\nu},
\end{equation}
with vacuum expectation,
\begin{equation}\label{p7}
\langle\theta^\mu_\mu\rangle=-4|\varepsilon_v|.
\end{equation}
The leading term of the QCD $\beta$-function is known to be, 
\begin{equation}\label{p8}
\beta=-\dfrac{11g^3}{(4\pi)^2}.
\end{equation}
Now, calculating the energy-momentum tensor trace of Eq.(\ref{tt7}) for the glueball field $\eta$ using the relation,
\begin{equation}\label{p9}
\theta^\mu_\mu=4V(\eta)+\eta\square\eta,
\end{equation}
we get, 
\begin{align}\label{p10}
\theta^\mu_\mu&=4V(\eta)-\eta\dfrac{\partial G}{\partial\eta}F^{\mu\nu}F_{\mu\nu}-\eta\dfrac{\partial V}{\partial\eta}\nonumber\\
&=4\tilde{V}-\eta G'(\eta)F^{\mu\nu}F_{\mu\nu},
\end{align}
where $G'$ and $V'$ represent first derivative with respect to $\eta$ and $\tilde{V}(\eta)=V(\eta)-\eta V'(\eta)/4$.  Also, rescaling $\tilde{V}(\eta)$ with the energy density $-|\varepsilon_v|$ i.e. $\tilde{V}(\eta)\rightarrow-|\varepsilon_v|\tilde{V}(\eta)$ together with the vacuum expectation value in Eq.(\ref{p7}) we get,
\begin{equation}\label{p11}
\left\langle\eta G'(\eta)F_{\mu\nu}F^{\mu\nu}\right\rangle=4|\varepsilon_v|\langle 1-\tilde{V}\rangle,
\end{equation}
with this equation, we recover the classical result in the limit $|\varepsilon_v|\rightarrow 0$ i.e. $\langle F_{\mu\nu}F^{\mu\nu}\rangle=0$. Using the potential expressed in Eq.(\ref{p2}) we can determine,
\begin{align}\label{p12}
\tilde{V}&=V-\dfrac{\eta V'}{4}\nonumber\\
&=\dfrac{m_\phi^2\eta^2}{4},
\end{align}
as a result, Eq.(\ref{p11}) can be expressed as
\begin{equation}\label{p13}
\left\langle 2G(\eta)F_{\mu\nu}F^{\mu\nu}\right\rangle=4|\varepsilon_v|\left\langle 1-\dfrac{m^2_\phi\eta^2}{4}\right\rangle,
\end{equation}
we can identify the gluon mass \cite{Issifu2}, 
\begin{equation}
m_A^2=\dfrac{m_\phi^2}{4}. 
\end{equation}
Furthermore, taking the expectation value of Eq.(\ref{cgt2}) in terms of the glueball field $\eta$, we can express
\begin{align}
\left\langle \dfrac{m^2_\phi\eta}{4}F^{\mu\nu}F_{\mu\nu}\right\rangle-\left\langle m^2_\phi\eta\right\rangle=0,
\end{align}
consequently, the mean glueball field $\bar{\eta}$ has two possible solution i.e. $\bar{\eta}=0$ and $\bar{\eta}=1$. Higher glueball condensate corresponds to $\bar{\eta}=0$ whilst lower glueball condensate corresponds to $\bar{\eta}=1$ \cite{Carter}.
\begin{figure}[H]
  \centerline{\includegraphics[scale=0.7]{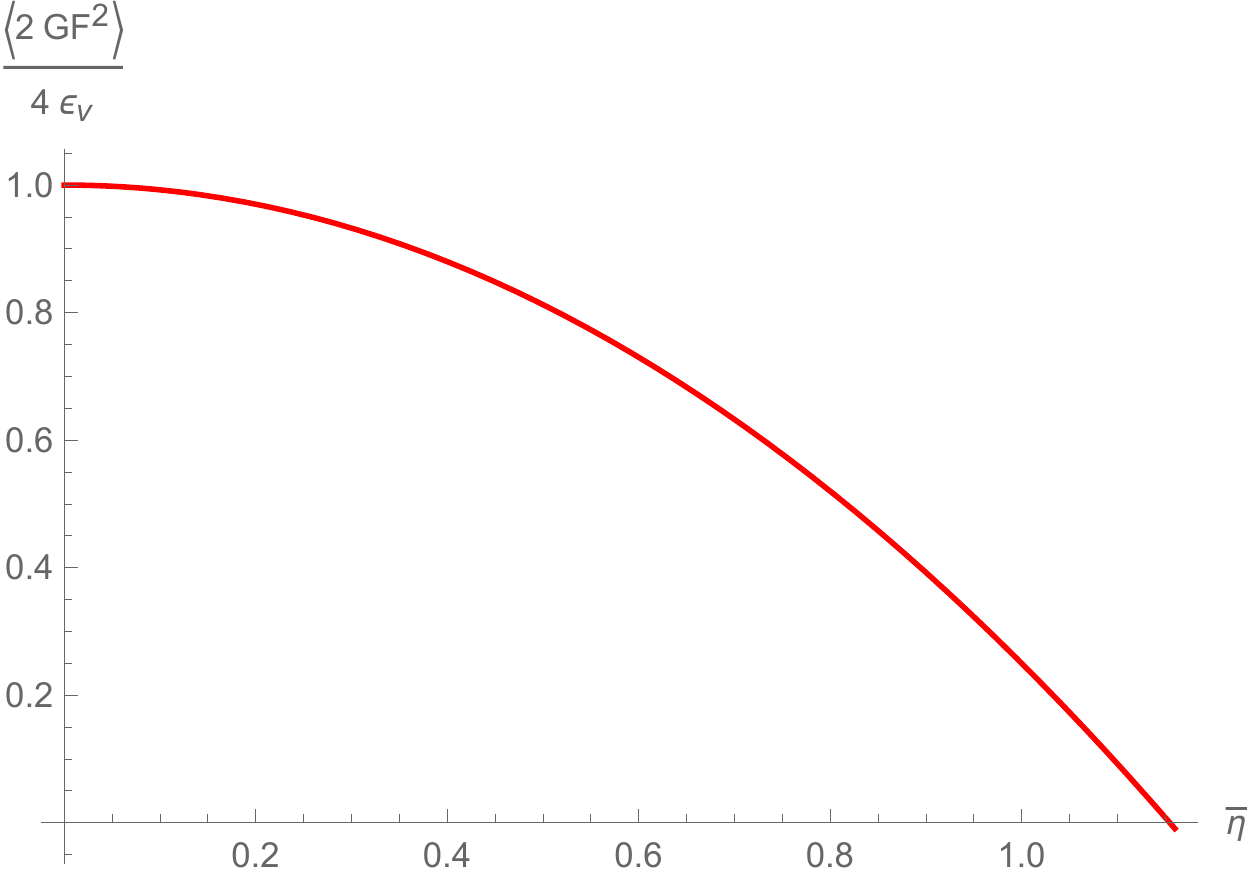}}
  \caption{Gluon condensation}
   \label{v1}
    \floatfoot{An increase in the mean glueball field $\bar{\eta}$ decreases the gluon condensate until it vanishes at maximum $\bar{\eta}$.}
\end{figure}

\subsection{Strong Running Coupling $\alpha_s$ and QCD $\beta$-Function}\label{SRC}
{Comparing  Eqs.(\ref{p7a}) and (\ref{p10}), we can relate 
\begin{align}
\dfrac{\beta(g)}{2g}=-\eta G'(\eta)\rightarrow\nonumber\\
\beta(1/r^2)=-2g\eta G'(\eta).
\end{align}
We can also extract the strong running coupling $\alpha_s$ using the renormalization group theory \cite{Deur},
\begin{align}\label{rn1}
\beta(Q^2)=Q^2\dfrac{d\alpha_s}{dQ^2},
\end{align}
comparatively;
\begin{equation}
\beta(\eta)\simeq-\eta\dfrac{d(G)}{d\eta}=-m_\phi^2\eta^2(r)=\beta(1/r^2) \qquad\text{we set}\qquad g=1.
\end{equation} %with the results obtained from the classical QCD Lagrangian,
%\begin{equation}\label{vg3c1}
%\theta^\mu_\mu=\sum\limits_f m_fq_f\bar{q}_f-\dfrac{b\alpha_s}{8\pi}F^{a\mu\nu}F^a_{\mu\nu},
%\end{equation}
%where $\beta(g)=-b\alpha_s/(4\pi)$ ($b=11$ depicts pure gluodynamics), since the theory do not contain fermions $m_f=0$ and $-\beta(1/r^2)=4G(\eta)$. Noticing that $4G(\eta)=2\eta G'(\eta)$, we can use the renomalization group theory
%\begin{equation}\label{p11a}
%\beta(Q^2)=Q^2\dfrac{d\alpha_s(Q^2)}{dQ^2},
%\end{equation}
%we can express
Therefore, the strong coupling can be identified as $\alpha_s(\eta)=G(\eta)=\alpha_s(1/r^2)$. QCD $\beta$-function is naturally a negative quantity showing the asymptomatic freedom nature of the strong coupling. It also reveals the anti-screening behavior of the theory at higher energies. The strong running coupling, on the other hand, gives an insight into the growing precision of hadron scattering experiments at high energy limits. And at low energy limits, within the scale of hadron mass, it enhances understanding of hadron structure, color confinement, and hadronization. Now, substituting the solution of the glueball field $\eta$ and expanding it for $r\rightarrow 0 $ we obtain,
\begin{align}
\alpha_s(1/r^2)=\left[1-\dfrac{m_\phi^2r^2}{3}\right].
\end{align}
%%&=\left[1- \dfrac{m_\eta^2r^2}{3}\right] \quad\text{where}\quad \xi^2=\dfrac{1}{\rho a^2}
Also, we can associate the spacelike momentum $Q$ with $Q\equiv 1/r$, then
\begin{equation}
\alpha_s(Q^2)=\left[1-\dfrac{m_\phi^2}{3Q^2} \right].
\end{equation}
In terms of the four-vector momentum i.e. $Q^2\equiv -q^2$, the strong coupling becomes
\begin{equation}
\alpha_s(q^2)=\left[1+\dfrac{m_\phi^2}{3q^2} \right],
\end{equation}
and the $\beta$-function becomes,
\begin{equation}
\beta(q^2)=-2\left[1+\dfrac{m_\phi^2}{3q^2} \right].
\end{equation}
We observe that in the limit $q^2\rightarrow 0$ the strong coupling shows a singularity, generally referred to as the Landau singularity. It marks the failure of perturbative QCD. The singularity is attributed to the self-interacting gluons with hadron degrees of freedom in the IR regime leading to color confinement \cite{Olive}. In that case, the gluons dynamically acquire mass at $q^2\rightarrow 0$ i.e. $q^2\cong m_A^2$ \cite{Badalian, Yu} which increases the coupling infinitely. Hence, the singularity can be removed by fixing a freezing point \cite{Badalian, Badalian1} to the strong running coupling at $q^2\cong m_A^2$ i.e.,
\begin{equation}
\alpha_s(q^2)=\left[1+\dfrac{m_\phi^2}{3(q^2+m^2_A)} \right],
\end{equation}
and 
\begin{equation}
\beta(q^2)=-2\left[1+\dfrac{m_\phi^2}{3(q^2+m^2_A)} \right].
\end{equation}
Thus, the 'so called' gluon mass is more pronounced at $q^2\rightarrow 0$ and its effect gradually fades off in the limit where $q^2\rightarrow\infty$ \cite{Cornwall1}. A recent analysis of this subject is contained in \cite{Deur}.
}

%\begin{equation}\label{p11b}
%\beta(1/r^2)=-\eta\dfrac{d}{d\eta}\left(2G(\eta)\right), 
%\end{equation}
%consequently, we can identify $\alpha(1/r^2)=2G(\eta)$, where $Q\equiv 1/r$. So, using the solution at the right side of Eq.(\ref{cgt12}) and expanding it to the quadratic term
%\begin{align}\label{p11c}
%\alpha_s(1/r^2)&=2\left[1-\dfrac{m^2_\phi r^2}{3} \right] \rightarrow\nonumber\\
%\alpha_s(Q^2)&=2\left[1-\dfrac{m^2_\phi}{3 Q^2} \right] 
%\end{align}
%noticing that $Q^2=-q^2$ is a space-like momentum and $q$ is the four-vector momentum hence,
%\begin{equation}\label{p11d}
%\alpha_s(q^2)=2\left[1+ \dfrac{m^2_\phi}{3 q^2} \right]
%\end{equation}
%so the $\beta$-function becomes
%\begin{equation}\beta(q^2)=-4\left[1+ \dfrac{m^2_\phi}{3 q^2} \right].
%\end{equation}

\begin{figure}[H]
  \centering
  \subfloat[Left Panel]{\includegraphics[scale=0.6]{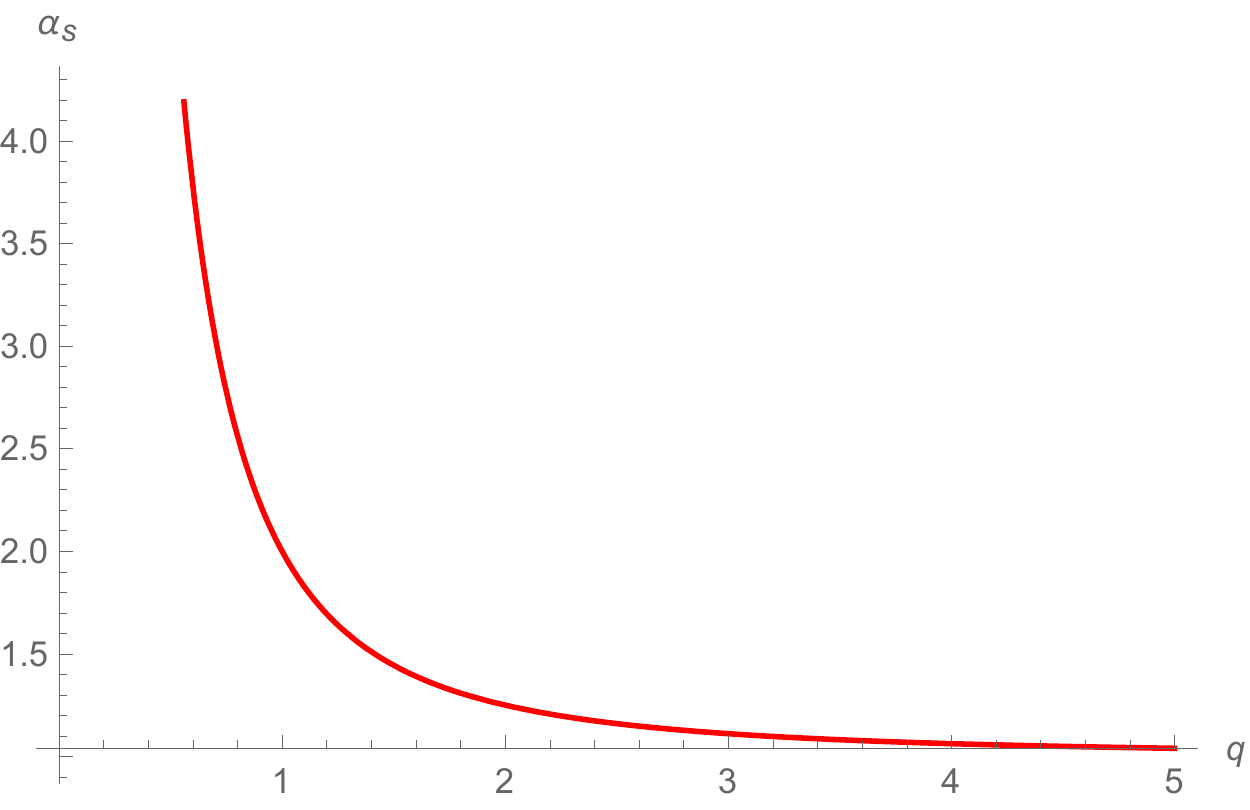}}
  \qquad
  \subfloat[Right Panel]{\includegraphics[scale=0.6]{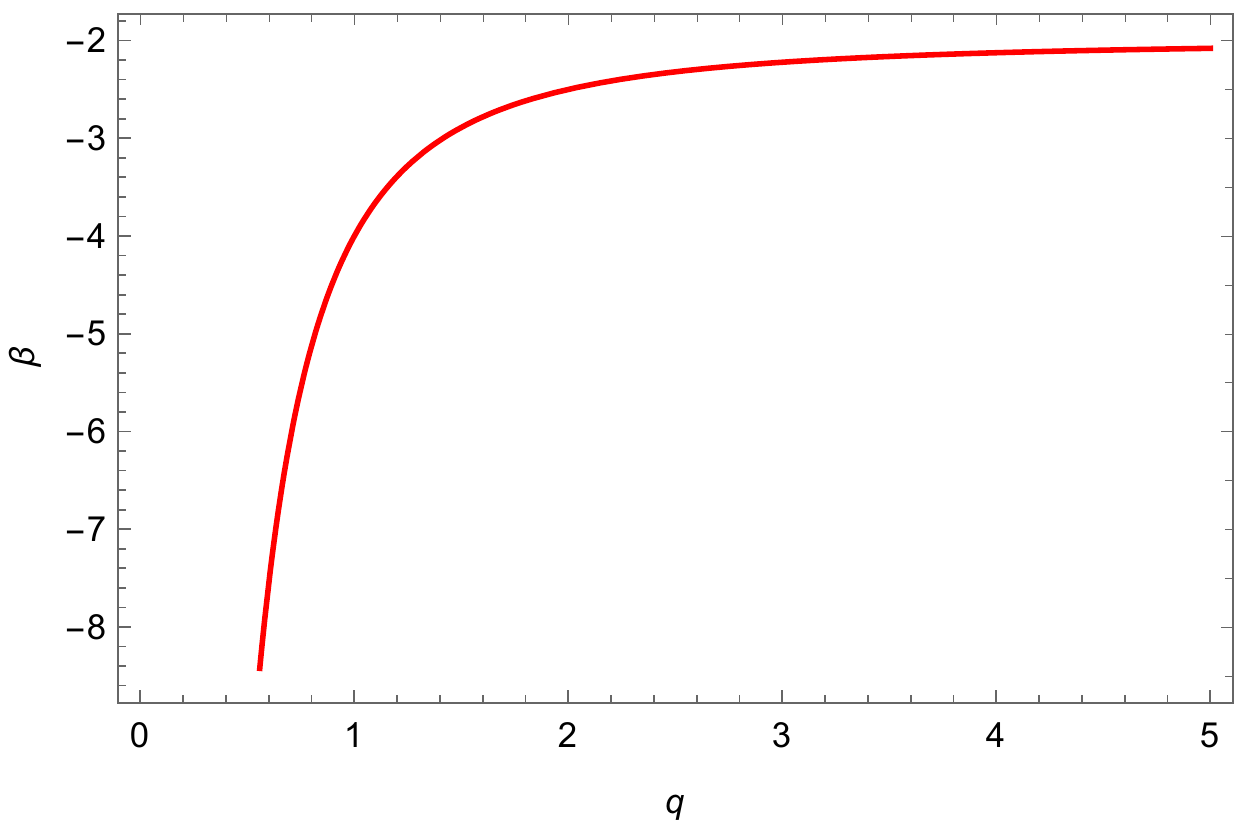}}
  \caption{Strong Running Couple $\alpha_s$ (left) and $\beta$-Function (right) against $q$, with a Landau Ghost Pole}
   \label{st1}
   \floatfoot{The graphs show an unphysical behavior at $q\rightarrow 0$, this is due to the presence of dynamically generated gluon mass. The self-interacting gluons and the strong force that exist between them are capable of creating bound states with hadron degrees of freedom. These graphs depict the behavior of $\alpha_s$ and $\beta$ observed from pQCD.}
\end{figure}

\begin{figure}[H]
  \centering
  \subfloat[Left Panel]{\includegraphics[scale=0.6]{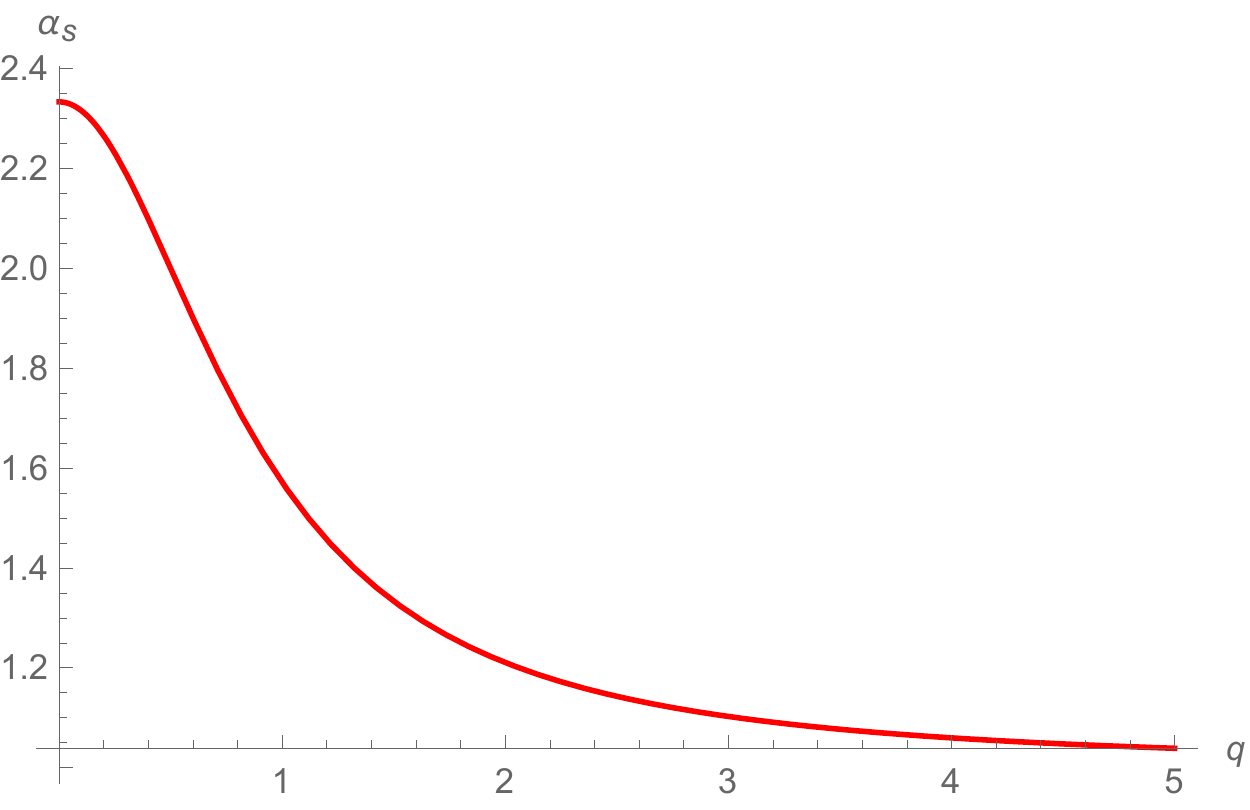}}
  \qquad
  \subfloat[Right Panel]{\includegraphics[scale=0.6]{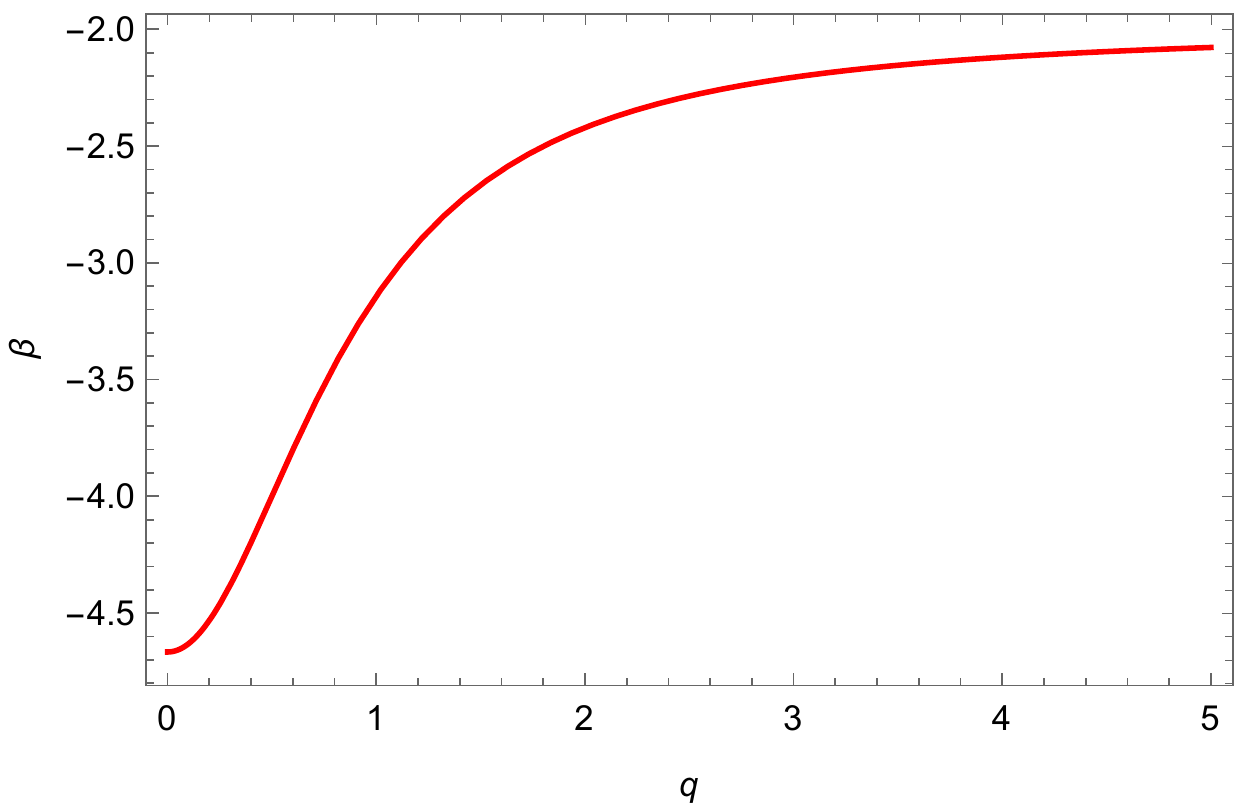}}
  \caption{Strong Running Coupling $\alpha_s$ (left) and $\beta$-Function (right) against $q$ with a freezing point at $q\rightarrow 0$}
   \label{st2}
   \floatfoot{Here, the Landau singularity has been fixed by introducing the gluon mass $m_A$. The presence of the gluon mass is more pronounced at $q\rightarrow 0$ and gradually vanishes in the limit $q\rightarrow \infty$. Consequently, $\alpha(0)\simeq 2.3$ and $\beta(0)\simeq-4.7$ from the graphs.}
\end{figure}
\section{Conclusion}\label{C}
We modified the DBI action to develop models that are capable of mimicking the phenomenon of QCD theory using an Abelian gauge field. The models were based on the behavior of opened string with its endpoints on the D$p$-brane. %At the tachyonic vacuum, the string on the brane behaves like a closed string because its endpoints gets connected by a flux line on the brane. Also, at the tachyonic vacuum the effective tachyon potential density cancels the brane tension, so it does not cost any energy to rearrange the string on the brane. At this point, the brane dissolves and its remnants behave as low-dimensional point particles at the endpoints of the string. 
In studying color particles, the endpoints of the string serve as the source and sink of the color charges. Additionally, the models are efficient in investigating glueballs when the tachyons condense and transform into glueballs with real square masses that keep them confined. Without fermions, the models are suitable for studying the bound states of gluons and the dynamics of glueballs.  To study the dynamics of quarks, we showed how the model can be coupled with standard model fermions systematically. Here, the particles involved are glueball-fermion-mix in a confined state. Moreover, we demonstrated that the color dielectric function coupled with the Abelian gauge capable of causing color confinement vanishes automatically when we introduced the non-Abelian gauge field in Sec.~\ref{NAGT}. Consequently, the presence of $G(\phi)$ coupled with the Abelian gauge field was to induce non-Abelian characteristics. 

We also developed one of the models to demonstrate its ability to explain some basic characteristics of strong interactions. We derived the linear and Cornell-like potentials that are used to describe color particles in phenomenological QCD. The linear potential is motivated by the string model of hadrons whilst the Cornell potential is motivated by Lattice QCD calculations. The Cornell potential is particularly important in QCD because it shows both the asymptotic freedom and color-confining behavior exhibited by the model. We also calculated the strong running $\alpha_s$ coupling and the QCD $\beta$-function and compared their behavior with the traditional QCD theory. However, in the model framework, we are able to fix the non-physical Landau ghost pole that occurs at the low energy region of the model by assuming the existence of gluon mass $m_A$ at low momentum region, $q^2\sim m^2_A$. The model leads to the determination of gluon condensate and how the glueball fields contribute to the condensate.

Furthermore, the models can be discretized using path integral formalism and investigated under lattice field theory with the availability of the required computational artifacts. As observed in the model developed in Sec.~\ref{TM}, the linear and the Cornell-like potentials can be used to study hadron and quarkonia spectra. Other hadron properties can also be studied from these models when the appropriate spin contributions to the potential are added. Extending the models to study the characteristics of particles at a finite temperature will pave way for understanding, chiral symmetry breaking and restoration, confinement/deconfinement, and quark-gluon-plasma phase transitions. Finally, the models can be applied in investigating physical systems such as pions. %In case of the real scalar field $\phi$, it can be used to investigate $\pi^+,\,\pi^0,\,\text{and},\,\pi^-$ systems with tachyons in their spectra, so
%\begin{equation}\label{qt21a1.}
%\phi\equiv\begin{pmatrix}\phi_1 \\ \phi_2\\\phi_3\end{pmatrix},
%\end{equation}
%which will be invariant under non-Abelian $\text{SU}(2)\sim\text{SO}(3)$ group representation,
%\begin{equation}
%\phi\rightarrow \phi'=e^{i\alpha^at_a}\phi.
%\end{equation}
%and transforms as $\text{SU}(2)\sim\text{SO}(3)$
%On the other hand, the charged scalar field,
%\begin{equation}\label{qt21a1..}
%\phi\equiv\dfrac{1}{\sqrt{2}}\begin{pmatrix}\phi_1 \\ i\phi_2\end{pmatrix},
%\end{equation}
%can be used to investigate charged pions $\pi^\pm$ systems with tachyons in their spectra. This can be represented under the rotation group,
% $\text{U}(1)\sim\text{SO}(2)$,
%\begin{equation}
%\phi=\begin{pmatrix}\phi_1 \\ \phi_2\end{pmatrix}\rightarrow\phi'=\left( {\begin{array}{cc}
 % \cos\theta & -\sin\theta \\
 %  \sin\theta & \cos\theta
 % \end{array} } \right)\begin{pmatrix}\phi_1 \\ \phi_2\end{pmatrix}.
%\end{equation}
% c + 0 / ) = 
\acknowledgments

This work was supported by Conselho Nacional de Desenvolvimento Cient\'ifico e Tecnol\'ogico (CNPq) project No.: 168546/2021-3, Brazil.
F.A.B. would like to thank CNPq, CAPES and CNPq/PRONEX/FAPESQ-PB (Grant No. 165/2018), for partial financial support. F.A.B. also acknowledges support from CNPq (Grant No. 312104/2018-9), Brazil.

\end{document}